\documentclass[]{aa}
\usepackage{natbib,twoopt}
\usepackage[breaklinks=true]{hyperref}
\usepackage{multirow,threeparttable}
\usepackage{txfonts}
\usepackage{color}
\usepackage{xcolor}
\usepackage{mathtools}
\usepackage{url}
\usepackage[nolist,nohyperlinks]{acronym}

\acrodef{BCG}[BCG]{brightest cluster galaxy}
\acrodefplural{BCG}[BCGs]{brightest cluster galaxies}
\acrodef{ICM}[ICM]{intracluster medium}
\acrodef{SZ}[SZ]{Sunyaev-Zel'dovich}
\acrodef{DSA}[DSA]{diffusive shock acceleration}
\acrodef{ACIS}[ACIS]{Advanced CCD Imaging Spectrometer}
\acrodef{CIAO}[CIAO]{Chandra Interactive Analysis of Observations}
\acrodef{CASA}[CASA]{Common Astronomy Software Applications}
\acrodef{GMRT}[GMRT]{Giant Metrewave Radio Telescope}
\acrodef{VLA}[VLA]{Very Large Array}
\acrodef{JVLA}[JVLA]{Jansky Very Large Array}
\acrodef{RFI}[RFI]{radio frequency interference}
\acrodef{NXB}[NXB]{non X-ray background}
\acrodef{APEC}[APEC]{Astrophysical Plasma Emission Code}
\acrodef{ARF}[ARF]{Ancillary Response File}
\acrodef{RMF}[RMF]{Response Matrix File}
\acrodef{LHB}[LHB]{local hot bubble}
\acrodef{GH}[GH]{galactic halo}
\acrodef{CXB}[CXB]{cosmic X-ray background}
\acrodef{CIE}[CIE]{collisional ionization equilibrium}
\acrodef{HD}[HD]{hydrodynamic}
\acrodef{MHD}[MHD]{magnetohydrodynamic}
\acrodef{KHI}[KHI]{Kelvin-Helmholtz instability}

\defcitealias{2014MNRAS.443.2463O}{O14}

\begin{document}

\title{Deep \emph{Chandra} observations of merging galaxy cluster ZwCl 2341+0000}
\author{X. Zhang\inst{1,2} 
\and A. Simionescu\inst{2,1,3} 
\and C. Stuardi\inst{4,5}
\and R. J. van Weeren\inst{1}
\and H. T. Intema\inst{1}
\and H. Akamatsu\inst{2} 
\and J. de Plaa\inst{2}
\and J. S. Kaastra\inst{2,1}
\and A. Bonafede\inst{4,5}
\and M. Br\"uggen\inst{6}
\and J. ZuHone\inst{7}
\and Y. Ichinohe\inst{8}
}
\institute{Leiden Observatory, Leiden University, PO Box 9513, 2300 RA Leiden, The Netherlands \\\email{xyzhang@strw.leidenuniv.nl}\label{inst1}
\and SRON Netherlands Institute for Space Research, Niels Bohrweg 4, 2333 CA Leiden, The Netherlands\label{inst2}
\and Kavli Institute for the Physics and Mathematics of the Universe (WPI), The University of Tokyo, Kashiwa, Chiba 277-8583, Japan\label{inst3}
\and Dipartimento di Fisica e Astronomia, Universit\`{a} di Bologna, via Gobetti 93/2, 40122 Bologna, Italy
\and INAF - Istituto di Radioastronomia di Bologna, Via Gobetti 101, 40129 Bologna, Italy
\and University of Hamburg, Hamburger Sternwarte, Gojenbergsweg 112, 21029 Hamburg, Germany
\and Harvard-Smithsonian Center for Astrophysics, 60 Garden Street, Cambridge, MA 02138, USA
\and Department of Physics, Rikkyo University, 3-34-1 Nishi-Ikebukuro, Toshima-ku, Tokyo 171-8501, Japan
}
\abstract
{Knowledge of X-ray shock and radio relic connection in merging galaxy clusters has been greatly extended in terms of both observation and theory over the last decade. ZwCl 2341+0000 is a double-relic merging galaxy cluster; previous studies have shown that half of the southern relic is associated with an X-ray surface brightness discontinuity, while the other half not. The discontinuity was believed to be a shock front. Therefore, it is a mysterious case of an only partial shock-relic connection.}
{By using the 206.5 ks deep \emph{Chandra} observations, we aim to investigate the nature of the southern surface brightness discontinuity. Meanwhile, we aim to explore new morphological and thermodynamical features. }
{We perform both imaging and spectroscopic analyses to investigate the morphological and thermodynamical properties of the cluster. In addition to the X-ray data, we utilize the GMRT 325 MHz image and JVLA 1.5 GHz and 3.0 GHz images to compute radio spectral index maps.  }
{
Surface brightness profile fitting and the temperature profile suggest that the previously reported southern surface brightness discontinuity is better described as a sharp change in slope or as a kink. This kink is likely contributed by the disrupted core of the southern subcluster.
The radio spectral index maps show spectral flattening at the south-eastern edge of the southern relic, suggesting that the location of the shock front is 640 kpc away from the kink, where the X-ray emission is too faint to detect a surface brightness discontinuity. We update the radio shock Mach number to be $\mathcal{M}_\mathrm{radio,S}=2.2\pm0.1$ and $\mathcal{M}_\mathrm{radio,N}=2.4\pm0.4$ for the southern and northern radio relics based on the injection spectral indices. We also put a $3\sigma$ lower limit on the X-ray Mach number of the southern shock to be $\mathcal{M}_\mathrm{X\text{-}ray,S}>1.6$. Meanwhile, the deep observations reveal that the northern subcluster is in a perfect cone shape, with a $\sim400$ kpc linear cold front on each side. This type of conic subcluster has been predicted by simulations but is observed here for the first time. It represents a transition stage between a blunt-body cold front and a slingshot cold front. Strikingly, we found a 400 kpc long gas trail attached to the apex of the cone, which could be due to the gas stripping. In addition, an over-pressured hot region is found in the south-western flank of the cluster. 
}
{}
\titlerunning{\emph{Chandra}  observations of ZwCl 2341+0000}
\authorrunning{X. Zhang et al.}

\maketitle

\section{Introduction}

As the nodes of the large-scale structures in the Universe, galaxy clusters grow hierarchically by accreting gas from cosmic filaments and merging with subclusters. Major galaxy cluster mergers are the most dramatic events in the Universe, releasing energies of up to $10^{64}$ erg. During the merger, the \ac{ICM}\ that belongs to each subcluster is mixed and shows a disturbed X-ray morphology. Two types of surface brightness jumps, shock fronts and contact discontinuities, are discovered in both observations \citep[e.g.,][]{2001ApJ...551..160V, 2002ApJ...567L..27M, 2009ApJ...704.1349O, 2010A&A...516A..32G, 2018MNRAS.476.5591B} and simulations \citep[e.g.,][]{2006ApJ...650..102A,2011ApJ...728...54Z} of merging clusters. 

Since the first report of the association of an X-ray shock and a radio relic \citep{2010ApJ...715.1143F} -- which is a type of morphologically elongated, spectrally steep, and polarized diffuse radio source observed in galaxy cluster peripheries -- dozens of radio relics have been confirmed to be associated with shock fronts \citep[see the review of][]{2019SSRv..215...16V}. The Mach number of shock fronts can be estimated either by radio observations based on the assumption of the \ac{DSA} theory \citep[e.g.,][]{1977DoSSR.234.1306K,1978MNRAS.182..147B,1978ApJ...221L..29B} or by X-ray observations based on the Rankine-Hugoniot condition \citep{1959flme.book.....L}. 
The Mach numbers estimated from radio observations are systematically higher than those based on X-ray observations \citep{2019SSRv..215...16V}, which could be due to the projection effects \citep{2015ApJ...812...49H} or the propagation of shocks through a magnetized turbulent ICM \citep{2021MNRAS.500..795D}.
Nevertheless, most of the merging shocks are in low Mach numbers ($\mathcal{M}\lesssim3$), which challenge the acceleration efficiency of the current \ac{DSA} theory. Therefore, galaxy clusters are unique laboratories for investigating the particle acceleration of weak shocks.


In contrast to shock fronts, contact discontinuities, also known as cold fronts, are sharp surface brightness and temperature edges that hold pressure equilibrium. The cold fronts in merging clusters are usually ``remnant core'' cold fronts \citep{2005ApJ...618..227T}, that is, the cores of the original subclusters, and are undergoing ram-pressure stripping \citep{1972ApJ...176....1G}. Typically, the remnant cores are in the shape of blunt bodies, for example, Bullet Cluster \citep{2002ApJ...567L..27M} and Abell 2146 \citep{2010MNRAS.406.1721R}.
By studying the cold fronts with the subarcsecond spatial resolution of \emph{Chandra}, knowledge of transport processes and the effect of magnetic fields in the \ac{ICM} has been well developed \citep[][]{2007PhR...443....1M,2016JPlPh..82c5301Z}.

ZwCl 2341+0000 ($z=0.27$) is a merging galaxy cluster that hosts radio relics in its southern and northern peripheries \citep{2009A&A...506.1083V}. The presence of the double relics usually suggests a binary merger with a simple merging geometry and a small angle between the merging axis and the plane of the sky. This merging axis angle was later confirmed to be $10\degr\substack{+34\\-6}$ by \citet{2017ApJ...841....7B}. The diffuse radio emission of ZwCl 2341+0000 was first reported by \citet{2002NewA....7..249B} in the 327 MHz \ac{VLA} image and 1.4 GHz NRAO VLA Sky Survey (NVSS) image. The two relics (denoted RN and RS) were later resolved in \ac{GMRT} 610 MHz, 241 MHz, and 157 MHz observations, where the radio halo is absent \citep{2009A&A...506.1083V}. The southern relic consists of three different components (RS1--3). 
By using a \ac{VLA} 1.4 GHz D-configuration observation, \citet{2010A&A...511L...5G} discovered large-scale filamentary diffuse emission along the entire cluster after removing point sources, which implies the possible presence of a radio halo. 
In optical frequencies, the galaxy distribution is NW-SE elongated. The dynamical mass of the system was first estimated to be $\sim1\text{--}3\times10^{15}\ M_\sun$ \citep{2013MNRAS.434..772B}. Later, \citet{2017ApJ...841....7B} pointed out that the dynamical mass is biased in this complex merging system, and, by using weak lensing, they reported a mass estimation of $(5.57\pm2.47)\times10^{14}\ M_\sun$, which is consistent with the mass estimation made using the \ac{SZ} effect by \emph{Planck}, $M_\mathrm{SZ}=(5.2\pm0.4)\times10^{14}\ M_\sun$ \citep{2016A&A...594A..27P}. Moreover, the optical analysis favors a three-subcluster model by using the Gaussian mixture model method, where a third subcluster with a different  redshift ($z=0.27432$) is spatially overlaid on the northern subcluster. The northern and southern subclusters are at redshifts of 0.26866 and 0.26844, respectively \citep{2017ApJ...841....7B}.

\citet[][]{2014MNRAS.443.2463O} studied the radio-X-ray connection at the radio relics using \emph{Chandra} and \emph{XMM-Newton} observations. They detected X-ray surface brightness discontinuities at both the southern and northern radio relics. The discontinuities were believed to be shock fronts, although spectral analysis was not possible due to the shallow observations. \citet{2014MNRAS.443.2463O} also found that only half of the southern radio relic is associated with X-ray surface brightness discontinuities. This was the original intention of obtaining the additional observations that are presented in this paper. The deep \emph{Chandra} observations not only unravel the previous mystery but also reveal new striking morphological features, for example, a cone-shaped subcluster that is predicted by simulations but had never before been observed. These new discoveries are explicitly described and discussed in the subsequent sections.

In this paper we present the results of the deep \emph{Chandra} observations of ZwCl 2341+0000. In addition, we utilized radio observations performed with the \ac{GMRT} at 325 MHz as well as the \ac{JVLA} L-band (1--2 GHz) B and C configurations and S-band (2--4 GHz) C and D configurations \citep[published by][]{2017ApJ...841....7B} to compute spectral index maps and probe the properties of the shock fronts. The organization of this paper is as follows. In Sect. \ref{sect:observation} we present the details of the X-ray and radio observations and data reduction. In Sect. \ref{sect:analysis} we describe the methods of data analysis. The radio spectral index map is given in Sect. \ref{sect:spx}. The detailed analysis results for each region of interest are presented in Sect. \ref{sect:result}. We discuss and conclude the work in Sects. \ref{sect:discussion} and \ref{sect:conclusion}.
We adopt a Lambda cold dark matter universe with cosmological parameters $\Omega_\Lambda=0.7$, $\Omega_0=0.3$, and $H_0=70$ km s$^{-1}$ Mpc$^{-1}$. At $z=0.27$, one arcminute corresponds to a physical size of 248 kpc.

\section{Observations and data reduction}\label{sect:observation}
\subsection{Chandra}
\begin{table*}
\caption{Details of the \emph{Chandra} observations.}
\begin{center}
\begin{tabular}{ccccccc}
\hline\hline
ObsID & Instrument & Mode & Pointing & Filtered Exposure & NXB scaling factor \tablefootmark{a} \\
&&& (RA$[\degr]$, Dec$[\degr]$, Roll$[\degr]$)& (ks) & \\
\hline
5786 & ACIS-I & VFAINT &(355.93, 0.29, 290) &26.6 & $0.909\pm0.016$\\\
17170 & ACIS-I & VFAINT & (355.91, 0.32, 291) & 23.5 & $0.889\pm0.008$\\
17490 & ACIS-I & VFAINT & (355.93, 0.28, 295) & 108.2 & $0.832\pm0.004$\\
18702 & ACIS-I & VFAINT & (355.91, 0.32, 291) & 23.4 & $0.881\pm0.017$\\
18703 & ACIS-I & VFAINT & (355.91, 0.32, 295) & 24.8 & $0.914\pm0.032$\\
\hline
\end{tabular}
\tablefoot{
\tablefoottext{a}{The scaling factor is defined as $f_\mathrm{obs}/f_\mathrm{bkg}$, where the $f_\mathrm{obs}$ and $f_\mathrm{bkg}$ are the 9--12 keV count rate in the observation set and in the stacked background set, respectively}
}
\end{center}
\label{tab:observation}
\end{table*}%

ZwCl 2341+0000 was observed by the \emph{Chandra} \ac{ACIS} five times (see Table \ref{tab:observation} for details). We used the \ac{CIAO} v4.12 \citep{2006SPIE.6270E..1VF} and CALDB v4.9.2 for data reduction. All level-1 event files were reprocessed by the task \texttt{chandra\_repro} with VFAINT mode background event filtering. Flare state events were filtered from the level-2 event files by using \texttt{lc\_clean}. The total filtered exposure time is 206.5 ks. 

We used stowed background files to estimate the non-X-ray background (NXB). All stowed background event files are combined and reprocessed by the task \texttt{acis\_process\_events} using the latest gain calibration files. The stacked stowed background of each ACIS-I chip has a total exposure time of 1.02 Ms.  For each ObsID, we scaled the NXB such that the 9-12 keV count rate matches the observation. The uncertainty of each scaling factor is calculated as the standard deviation of the scaling factors of individual ACIS-I chips, which is listed in the last column of Table \ref{tab:observation}.

\begin{figure*}[]
\begin{center}
\begin{tabular}{cc}
\includegraphics[width=0.45\hsize]{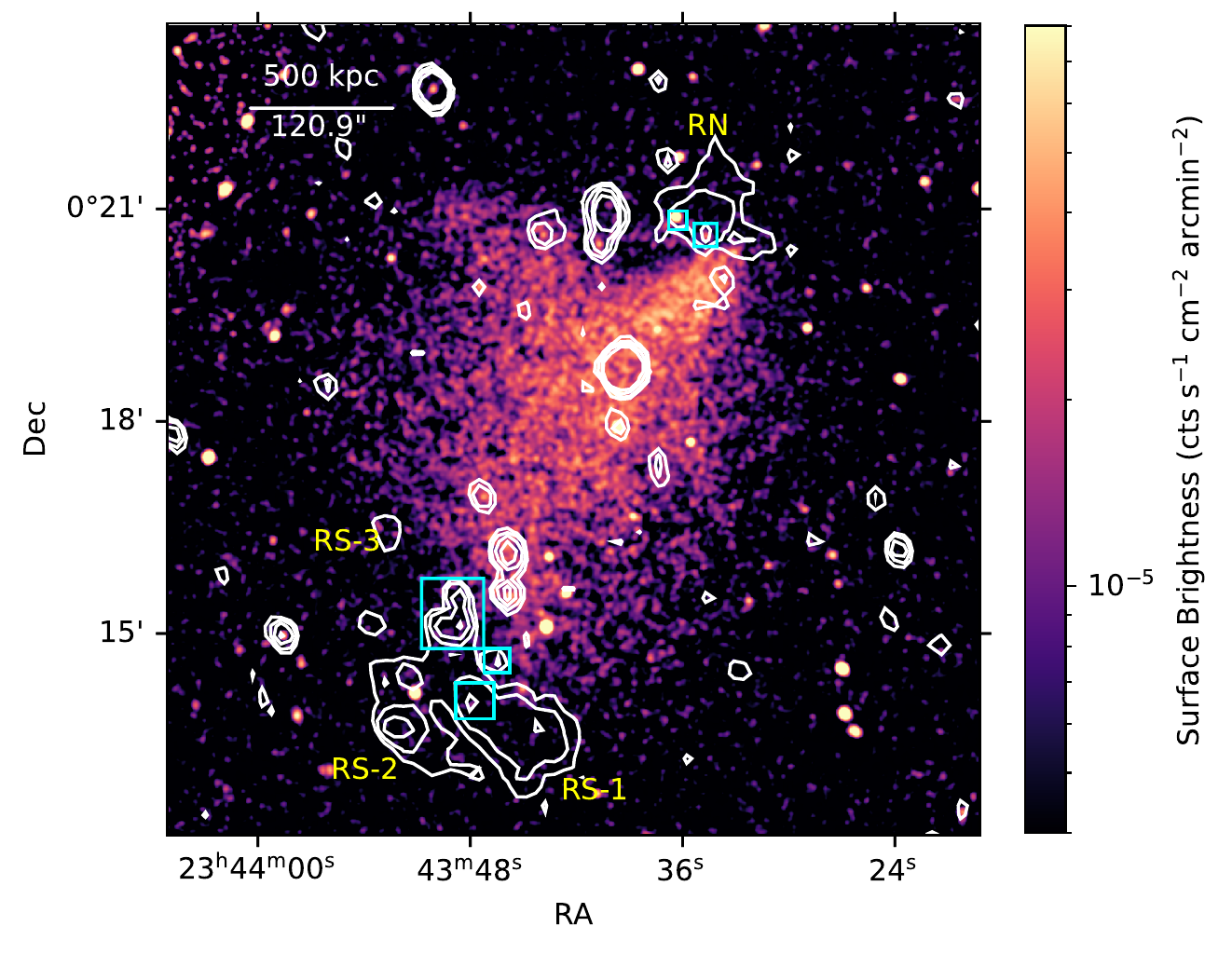}&
\includegraphics[width=0.45\hsize]{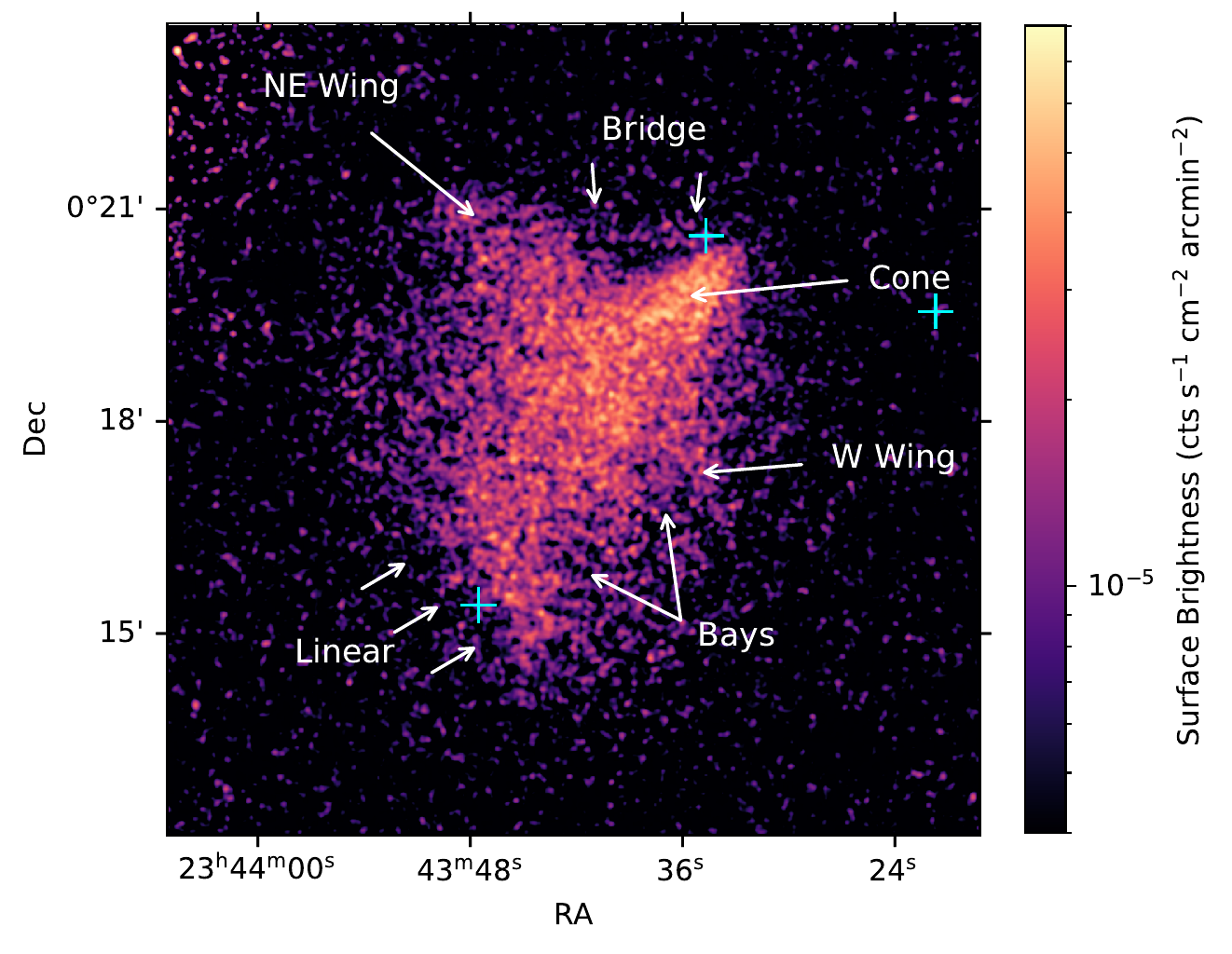}\\
\end{tabular}
\includegraphics[width=0.85\hsize]{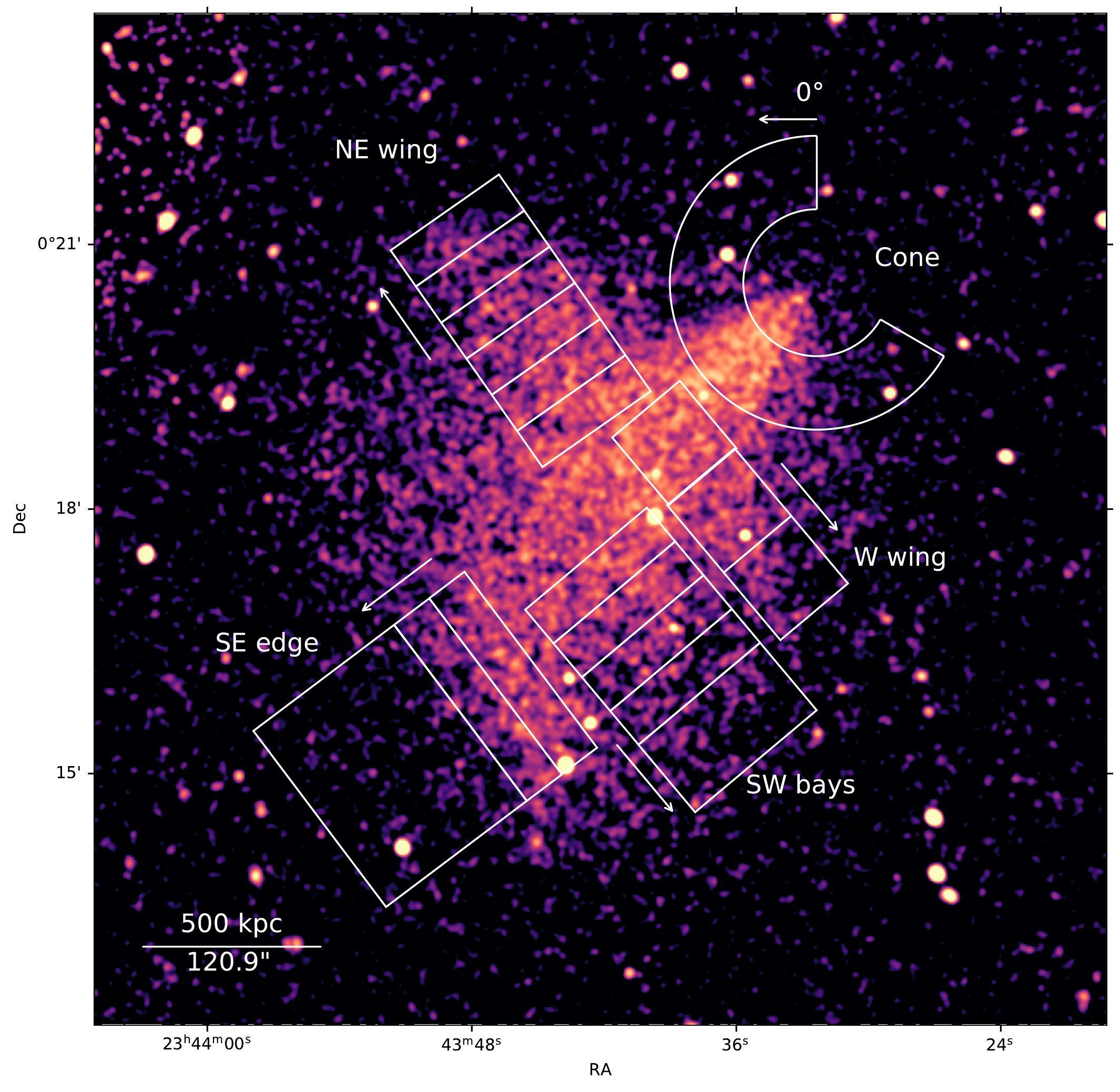}
\caption{
\emph{Top left:} X-ray flux map in the 0.5--2.0 keV band convolved with a $\sigma=1.97\arcsec$ (four pixel) Gaussian kernel. Overlaid are the \ac{JVLA} C-array 1.5 GHz radio contours at $[1, 2, 4]\times0.075$ mJy beam$^{-1}$. The terms of the relic sources RS-1--3 and RN are introduced by \citet{2009A&A...506.1083V}. Identified point sources that overlap with the southern relic are marked by cyan boxes. 
\emph{Top right:} Image from which point sources have been removed. Surface brightness features are labeled. The locations of the three brightest cluster galaxies adopted from \citet{2017ApJ...841....7B} are marked as cyan crosses. 
\emph{Bottom:} Extraction regions used for detailed X-ray analysis in Sect. \ref{sect:result}. For the regions SE edge, SW bays, western wing, and NE wing, the individual temperature bins in Fig. \ref{fig:pro} are plotted. The arrows beside each bin indicate the directions of the profiles.
}
\label{fig:image}
\end{center}
\end{figure*}

\subsection{GMRT and JVLA}

\begin{table*}
\caption{Details of the radio observations.}
\begin{center}
\begin{tabular}{cccccccc}
\hline\hline
Telescope & Project ID & Frequency & Band width & Configuration & Time on source &  Beam size & $\sigma_\mathrm{rms}$\\
&&(MHz)&(MHz)&&(hr)&($\arcsec$) & (mJy beam$^{-1}$) \\
\hline
GMRT & 20-061 & 325 & 33 & - & 22.67 & $10.7\times8.7$ & $7.0\times10^{-2}$ \\
\multirow{2}{*}{JVLA} & 17A-083 & 1500 & 1000 & C & 1.59 & $15.1\times11.1$ & $3.0\times10^{-2}$ \\
& SG0365 & 1500 & 1000 & B & 2.27 & $4.7\times3.4$& $1.2\times10^{-2}$\\
\hline
\end{tabular}
\end{center}
\label{tab:radio}
\end{table*}

\begin{figure*}[]
\begin{center}
\begin{tabular}{cc}
\includegraphics[height=0.37\hsize]{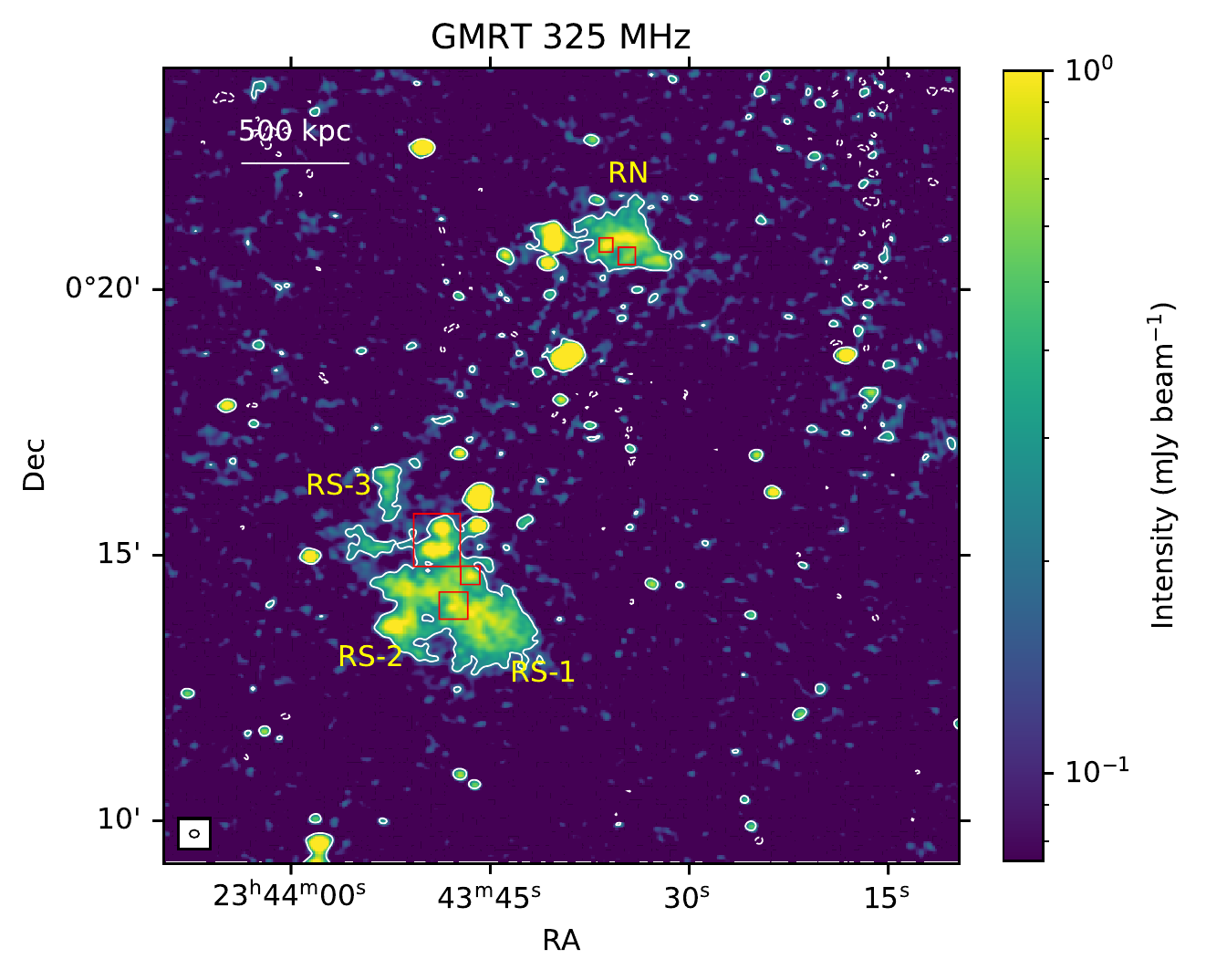}&
\includegraphics[height=0.37\hsize]{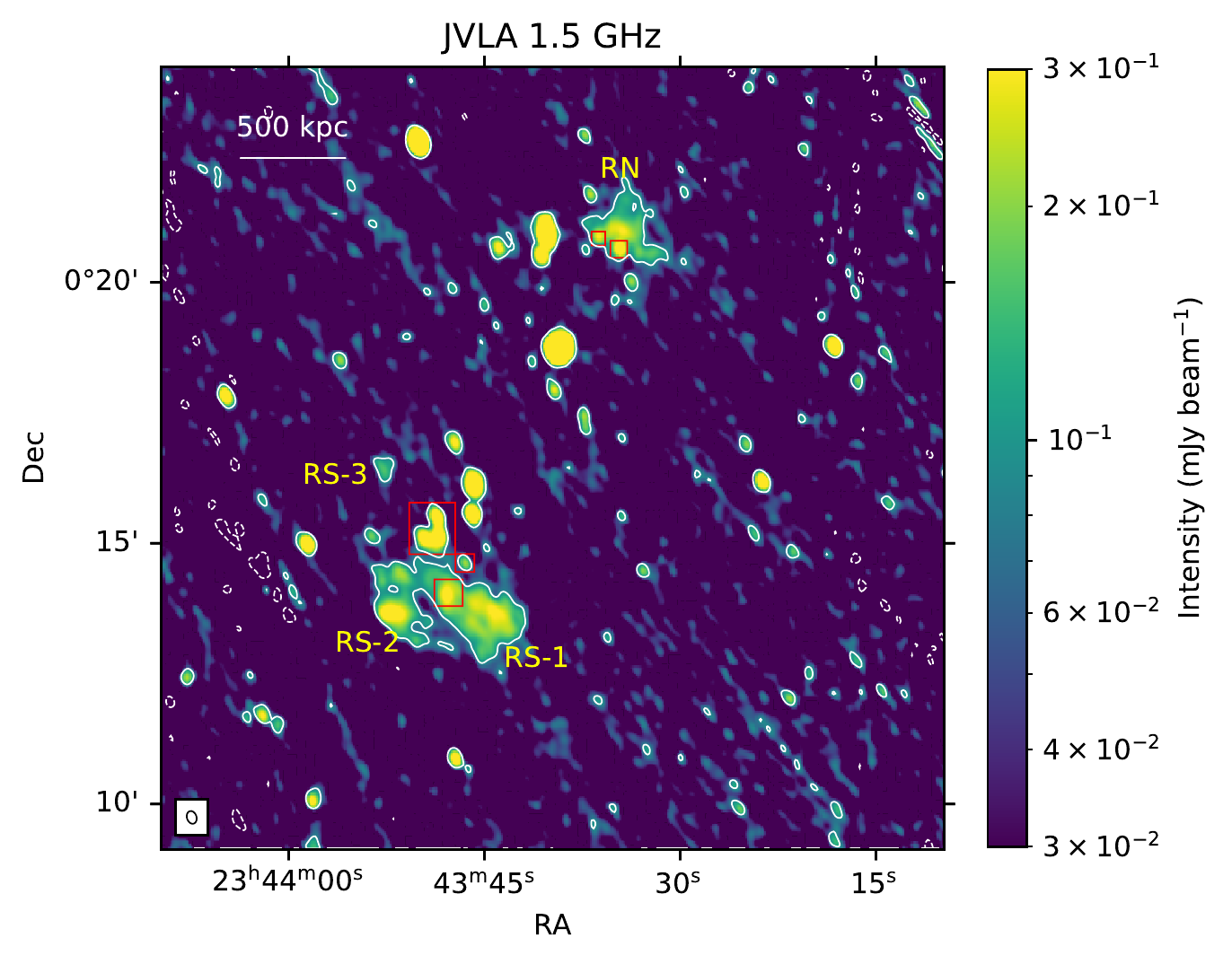}\\
\end{tabular}
\caption{
\ac{GMRT} 325 MHz (left) and \ac{JVLA} 1.5 GHz C-configuration (right) radio maps of ZwCl 2341+0000, overlaid with contours at the $3
\sigma_\mathrm{rms}$ (solid) and $-3\sigma_\mathrm{rms}$ (dashed) levels. The beam sizes are plotted at the bottom-left corner of each image. Identified point sources that overlap with the southern relic are marked by red boxes.
}
\label{fig:radio}
\end{center}
\end{figure*}

We use the \ac{GMRT} 325 MHz observations (project ID 20-061) and the \ac{JVLA} L-band observations (1-2 GHz band and 1.5 GHz central frequency) in B (project ID SG0365) and C (project ID 17A-083) configurations. The details of the observations are listed in Table \ref{tab:radio}. 

The \ac{GMRT} data were reduced and imaged by using the SPAM package \citep{2014ASInC..13..469I}, which is a set of AIPS-based data reduction scripts that includes ionospheric calibration \citep{2009A&A...501.1185I}. 
The JVLA data were reduced using the \ac{CASA} package 5.6.2. The B-configuration data were calibrated following the procedures of \citet{2016ApJ...817...98V}.
The C-configuration data were preprocessed by the \ac{VLA} CASA calibration pipeline, which performs basic flagging and calibration optimized for Stokes I continuum data. Then, we extracted uncalibrated data and we manually derived final delay, bandpass, gain, and phase calibration tables and applied them to the target. We used the source 3C138 as amplitude calibrator and we imposed the \citet{2013ApJS..204...19P} flux density scale. J0016-0015 was chosen as phase calibrator. We removed \ac{RFI}\ with CASA statistical flagging algorithms.

The two JVLA data sets were imaged using the multi-scale multifrequency deconvolution algorithm of the CASA task \texttt{tclean} \citep{2011A&A...532A..71R} for wide-band synthesis-imaging. We also used the w-projection algorithm to correct for the wide-field non-coplanar baseline effect \citep{2008ISTSP...2..647C}. We subtracted out from the visibilities all the sources external to the field of interest ($\sim15\arcmin\times15\arcmin$) to exclude bright sources that could increase the image noise. In order to refine the antenna-based phase gain variations, several cycles of self-calibration were performed. During the last self-calibration cycle, amplitude gains were also computed and applied. In the final images we used a Briggs weighting scheme with the robust parameter set to 0.5. The final images were corrected for the primary beam attenuation using the \texttt{widebandpbcor} task in CASA. The residual calibration errors on the amplitude are estimated to be $\sim5\%$.

The radio relics are visible in the GMRT 325 MHz image and the JVLA L-band C-configuration image (see Fig. \ref{fig:radio}), but are barely detected in the \ac{JVLA} B-configuration image. Therefore, we only use the B-configuration image for point source identification. In the GMRT 325 MHz image, the component RS-3 is more extended and another diffuse component emerges between the RS-3 and RS-2, suggesting the connection of all individual components.

\section{X-ray imaging and spectroscopy}\label{sect:analysis}

\subsection{Imaging analysis}
We extracted and combined the 0.5--2.0 keV count images using \texttt{merge\_obs}. The exposure maps are calculated using a weighted spectrum file generated by \texttt{make\_instmap\_weighted}, where the spectral model is a single \ac{APEC} model with $kT=5$ keV, which is the averaged temperature measured by \emph{Suzaku} \citep{2013PASJ...65...16A}. The \ac{NXB} count images were generated from the stowed observation files described in Sect. \ref{sect:observation}. After applying the combined exposure map to the \ac{NXB}-subtracted count image, we obtained the final X-ray flux map (see Fig. \ref{fig:image}). 

Point sources were detected from the unbinned count image by using the task \texttt{wavdetect} with wavelet scales of 1.0, 2.0, 4.0, 8.0, and 16.0 pixels. The parameter \texttt{sigthresh} was set to $10^{-6}$. The output region file is used for masking point sources in both imaging and spectral analysis. Meanwhile, we used the task \texttt{roi} to generate background regions around detected point sources and used the task \texttt{dmfilth} to create an image containing only the diffuse emission. We note that the point source subtracted image can better illustrate the surface brightness features, but the flux in point source regions is interpolated from the ambient background regions, which is not the exact value of the \ac{ICM} emission at the respective location. Therefore, in both imaging and spectroscopic analyses, we excluded the point sources using the region files from \texttt{wavdetect}.

We extracted surface brightness profiles and fitted them using the Sbfit\footnote{\url{https://github.com/xyzhang/sbfit}} v0.2.0 package \citep{xiaoyuan_zhang_2021_5507575}. The package uses a forward fitting strategy together with C-statistics \citep{1979ApJ...228..939C}.

\subsection{Spectral analysis}\label{sec:spec}

We used the task \texttt{specextract} to extract source and \ac{NXB} spectra and create the corresponding weighted ancillary response files (ARFs) and response matrix files (RMFs).
We used the spectral fitting package SPEX v3.06 \citep{1996uxsa.conf..411K,kaastra_j_s_2020_3939056} for spectral analysis. The original format spectra and response files were converted to SPEX format by the \texttt{trafo} task. We fit the spectra in the 0.5--7.0 keV band. All spectra were optimally binned \citep{2016A&A...587A.151K} and fitted with C-statistics \citep{1979ApJ...228..939C,2017A&A...605A..51K}. The reference proto-solar element abundance table is from \citet{LandoltBornstein2009:sm_lbs_978-3-540-88055-4_34}. By using Pyspextools \citep{jelle_de_plaa_2020_3707693}, the \ac{NXB} spectra were first smoothed by a Wiener filter with a window length of seven spectral channels and then subtracted from the source spectra. We fit the spectra of the five observations simultaneously. 

The source spectra are modeled using a spectral component combination of $cie1+(cie2+pow)\times hot+cie3\times red\times hot$, where $cie$ 1--3 are the single temperature \ac{CIE} models for the \ac{LHB}, \ac{GH}, and the \ac{ICM}; for $cie3$, the abundances of elements from Li to Mn are fixed to $0.3Z_\sun$, which is the averaged value of this cluster (see Sect. \ref{sec:result-general}). While for $cie$ 1 and 2, all abundances are fixed to the proto-solar abundance. The power law model for the \ac{CXB} residual is denoted as "pow" and its photon index is fixed to 1.41. The model $red$ denotes the redshift of the cluster, which is fixed to 0.27. The $hot$ is the \ac{CIE} absorption model, where the temperature is fixed to $5\times10^{-4}$ keV to represent the absorption from the Galactic neutral gas, and the column density is fixed to $n_\mathrm{H}=3.7\times10^{20}$ cm$^{-2}$, which is from the tool \texttt{nhtot}\footnote{\url{https://www.swift.ac.uk/analysis/nhtot/index.php}} \citep{2013MNRAS.431..394W}. Details of the modeling of \ac{LHB}, \ac{GH}, and \ac{CXB} residual are described in Sect. \ref{sec:xbg_model}.

\subsection{X-ray background modeling}\label{sec:xbg_model}

\begin{table*}
\caption{X-ray foreground and background components constrained by the RASS and \emph{Chandra} offset spectra. 
}
\label{tab:xbg}
\centering
\begin{tabular}{ccccc}
\hline\hline
Spectrum & Component & Flux \tablefootmark{a} ( 0.1 -- 2.4 keV) & $kT$ & $\Gamma$ \\
&& $10^{-2}$ ph s$^{-1}$ m$^{-2}$ & keV &\\
\hline
\multirow{2}{*}{RASS}&\ac{LHB} & $3.1\pm0.3$ & $0.10$ (fixed) & -\\
&\ac{GH} & $3.0\pm0.4$ & $0.151\pm0.007$ &-\\
\hline
\multirow{3}{*}{\emph{Chandra} offset}  & \ac{LHB} & $3.1$ (fixed)& $0.10$ (fixed) & -\\
&\ac{GH} & $2.4\pm1.3$ & $0.17\pm0.02$ &\\
&\ac{CXB} residual & $0.69\pm0.07$ & - & 1.41 (fixed)\\
\hline
\end{tabular}
\tablefoot{
\tablefoottext{a}{The normalizations are scaled to a 1 arcmin$^2$ area.}
}
\end{table*}

We used the \emph{ROSAT} All-Sky Survey (RASS) spectra as a reference to study the X-ray foreground components. The spectrum of an $0.3\degr$--$1.0\degr$ annulus centered at the cluster was extracted using the ROSAT X-Ray Background Tool \citep{2019ascl.soft04001S}. Similar to the source spectra, we fit the ROSAT background spectrum using a spectral model combination of $cie1+hot\times(cie2+pow)$. To solve the degeneracy of the \ac{LHB} and \ac{GH} temperatures, we fixed the \ac{LHB} temperature to a typical value of $0.1$ keV \citep[e.g.,][]{2000ApJ...543..195K}. The best-fit values are listed in Table \ref{tab:xbg}.

\begin{figure}
\begin{center}
\includegraphics[width=1\hsize]{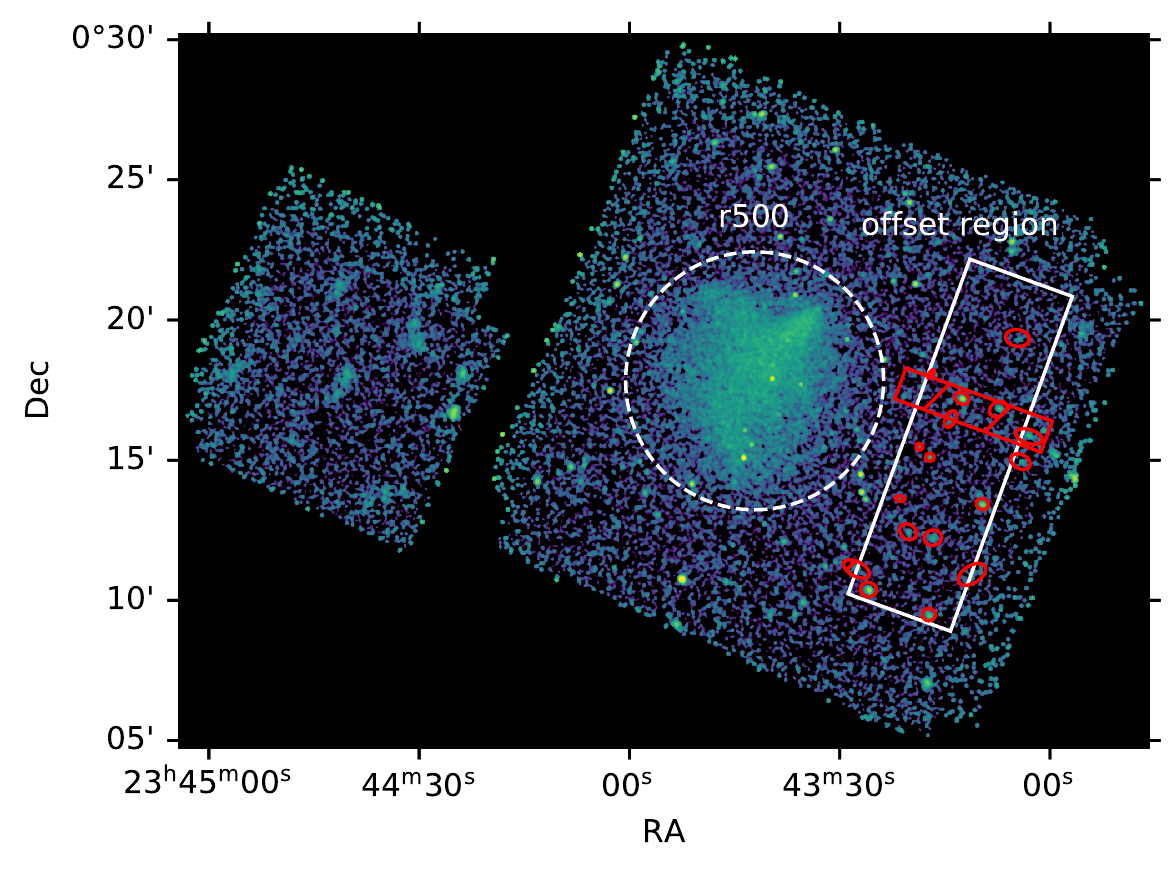}
\caption{Illustration of the offset extraction region, which is the white rectangle. Point-like sources are excluded using red regions, including the red rectangle, which is used to collectively exclude a series of observed clumps. The dashed circle is the $r_{500}$ region of the cluster, which is centered at [23:43:42.1,+00:17:49.9].}
\label{fig:offset}
\end{center}
\end{figure}

Meanwhile, we extracted spectra from an offset region outside the cluster's $r_{500}$ to cross-check the foreground component (see Fig. \ref{fig:offset}). The spectral components are the same as those of the RASS spectrum. Because the effective area of \emph{Chandra} drops dramatically below 1 keV, the \ac{LHB} component cannot be constrained. We fixed the temperature and normalization of the \ac{LHB} component to the values of the RASS spectrum. The best-fit results of the \emph{Chandra} offset spectra are listed in Table \ref{tab:xbg} as well. Though with large uncertainties, the normalization as well as the temperature of the \ac{GH} component are consistent to the values of the RASS spectrum fit. Therefore, in the source spectral analysis, we fixed the parameters of \ac{LHB} and \ac{GH} to the values of the RASS spectrum and fixed the parameters of the \ac{CXB} residual to the value of the \emph{Chandra} offset spectra.

\subsection{Systematic uncertainty estimation}
The \ac{GH} and \ac{CXB} residual components as well as the scaling factor of the \ac{NXB} can introduce uncertainties in the fit results. The best-fit normalization of the \ac{GH} in the \emph{Chandra} offset spectrum is $20\%$ different from the value of the RASS spectrum. Therefore, we use $20\%$ to estimate the uncertainty introduced by the \ac{GH} component. For the \ac{CXB} residual, converted to the 2--8 keV band, the surface brightness is $1.8\times10^{-15}$ erg s$^{-1}$ cm$^2$ arcmin$^{-2}$. Using Eqs. C.4 and C.5 in \citet{2020A&A...642A..89Z} and assuming the $\log N-\log S$ relation is identical to the \emph{Chandra} Deep Field South \citep{2012ApJ...752...46L}, the corresponding point source exclusion limit is $3\times10^{-15}$ erg s$^{-1}$ cm$^{-2}$ and the \ac{CXB} residual uncertainty on a 1 arcmin$^2$ area is $57\%$. For any of the spectra in the analysis, the \ac{CXB} residual uncertainty is scaled from $57\%$ by a factor of $A^{-1/2}$, where $A$ is the extraction area converted from the \texttt{backscal} value in the spectrum header. 
The largest NXB uncertainty amounts to $3\%$ in ObsID 18073; therefore, we conservatively used $3\%$ to estimate the NXB systematic uncertainty for all five observations.

\begin{figure*}[]
\begin{center}
\includegraphics[height=2.6 in]{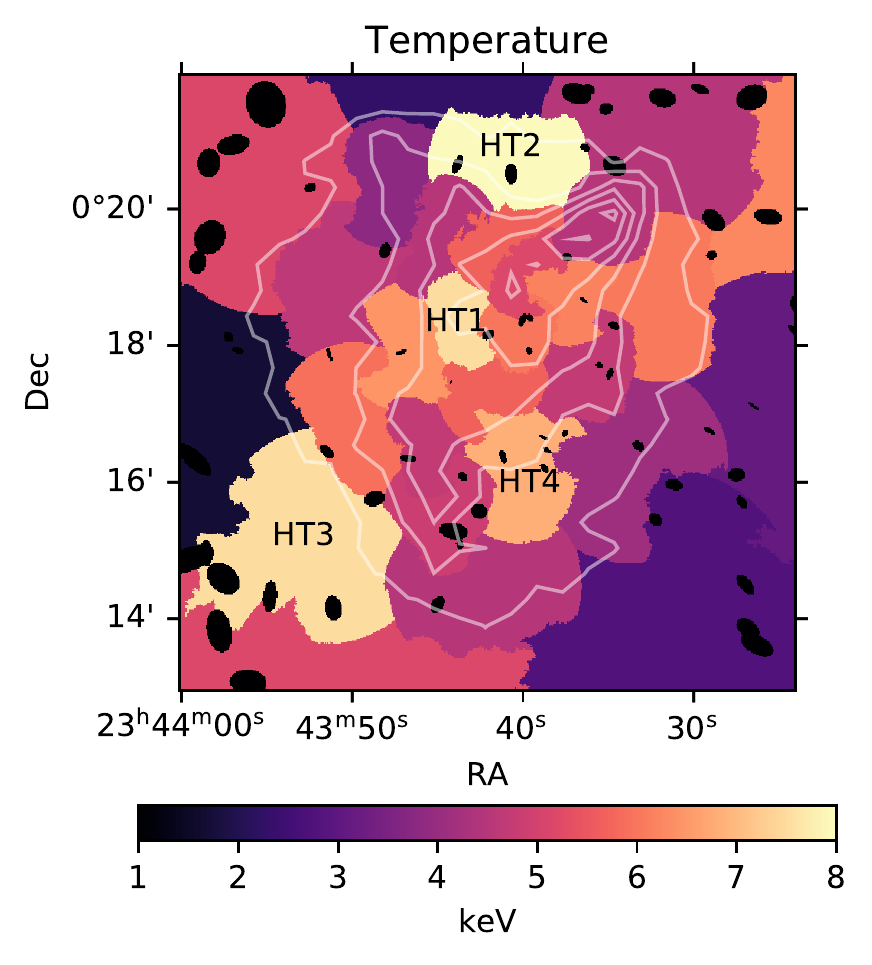}
\includegraphics[height=2.6 in]{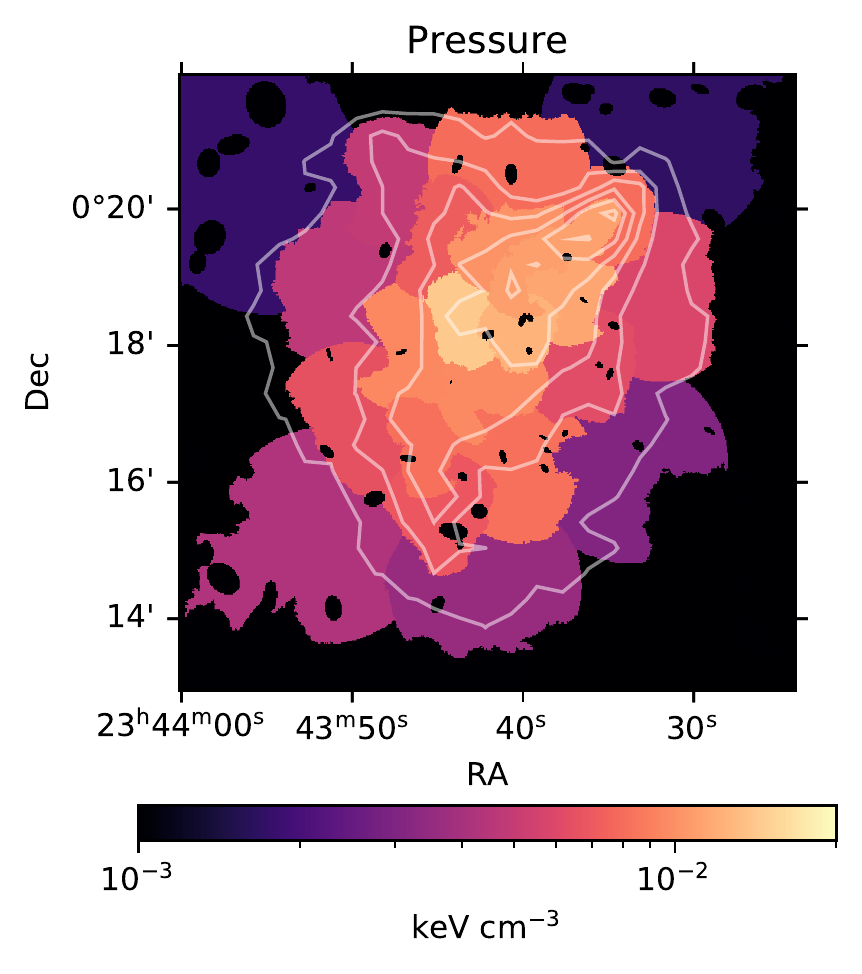}
\includegraphics[height=2.6 in]{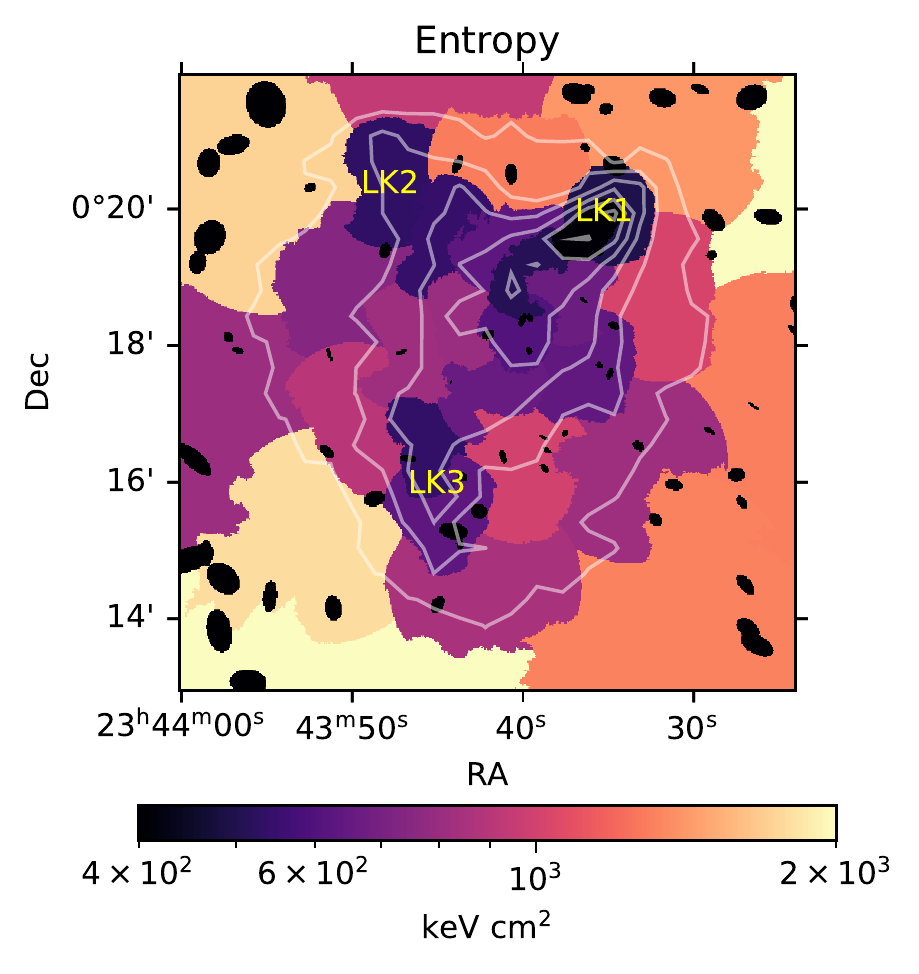}
\caption{Temperature, pseudo pressure, and entropy maps of ZwCl 2341+0000, overlaid with the X-ray surface brightness contours. High temperature and low entropy regions are labeled as HT1--4 and LK1--3, respectively.}
\label{fig:thermomap}
\end{center}
\end{figure*}

By fitting the spectra extracted inside $r_{500}$, which include most of the \ac{ICM} photons, we find that the statistical uncertainty is larger than the systematic uncertainties from the \ac{GH} and \ac{CXB} by an order of magnitude, and is larger than the NXB systematic uncertainty by four orders of magnitude. 
We also checked the systematic uncertainties in two smaller regions, the LK1 and HT3 regions in the thermodynamic maps (see Fig. \ref{fig:thermomap}), which are in a high and a low surface brightness, respectively. In the LK1 region, the statistical uncertainty is larger than all other systematic uncertainties by over an order of magnitude. In the HT3 region, the systematic uncertainties increase due to the low surface brightness of the ICM emission. Nevertheless, the statistical uncertainty is still a factor of four larger than the systematic uncertainty contributed by GH, and over an order of magnitude higher than other systematic uncertainties. The one-fourth of the additional uncertainty will result in a $3\%$ increase after uncertainty propagation. 
Therefore, in this work, we only present statistical uncertainties.

\subsection{Thermodynamic maps}
To create spatial bins for temperature, pseudo pressure, and entropy maps, we used the tool \texttt{contbin}\footnote{\url{https://github.com/jeremysanders/contbin}} \citep{2006MNRAS.371..829S}. The signal to noise ratio was set to 30 and the geometric constraint was set to 1.3. The spectra in each spatial bin were fitted following the method in Sect. \ref{sec:spec}. Assuming the depth along the line of sight of each bin is 1 Mpc and $n_e/n_\mathrm{H}=1.2$, we calculated the \ac{ICM} number density from the best-fit emission measure. The pseudo pressure is the product of the averaged gas density and the temperature of each bin. The entropy is calculated using the definition $K=kTn_e^{-2/3}$.

\begin{figure}
\begin{center}
\includegraphics[width=0.99\hsize]{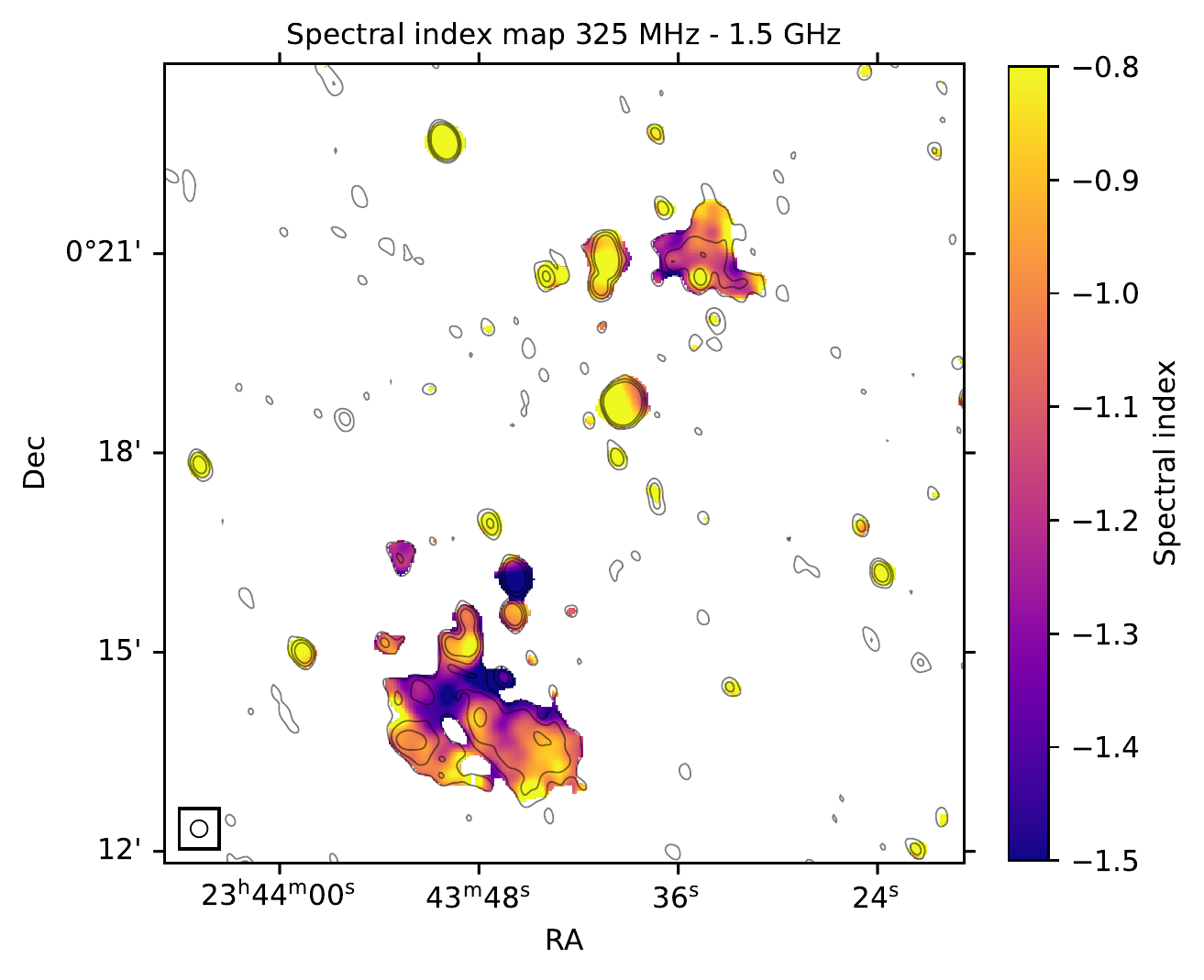}
\caption{Spectral index map between 325 MHz and 1.5 GHz with a resolution of $15.5\arcsec$, overlaid with \ac{JVLA} C-configuration contours. }
\label{fig:spx}
\end{center}
\end{figure}

\section{Radio spectral index map}\label{sect:spx}

Combining the \ac{GMRT} image at 325 MHz and the \ac{JVLA} C configuration image at 1.5 GHz we made a spectral index\footnote{$F\propto\nu^{\alpha}$} map of the cluster. We convolved the two images to the same Gaussian beam of 15.5 x 15.5 arcsec and excluded pixels below the respective $3\sigma_\mathrm{rms}$ detection threshold in each image. In our case, the JVLA observation is shallower; therefore, we excluded the pixels in the convolved GMRT image to match the convolved JVLA image. The spectral index image was created using the CASA \texttt{immath} task with \texttt{spix} mode. The error map image was created propagating the rms noise of each image and the calibration error with the expression:
\begin{equation}
\sigma_\alpha=\frac{1}{\log\left(\nu_1/\nu_2\right)}\times 
\left[\sigma_\mathrm{sys}^2+\left(\frac{\sigma_{\nu_1}}{I_{\nu_1}}\right)^2+\left(\frac{\sigma_{\nu_2}}{I_{\nu_2}}\right)^2\right]^{1/2},
\end{equation}
where $\nu_1$ and $\nu_2$ are the frequencies of the two images, $\sigma_{\nu}$ is the rms noise of the image and $I_{\nu}$ is the total intensity image at the frequency $\nu$, and  $\sigma_\mathrm{sys}$ is the calibration uncertainty, which is $5\%$ here. The spectral index map between 325 MHz and 1.5 GHz is shown in Fig. \ref{fig:spx}.

The spectral index at the SE edge of the radio relic RS-2 is -1.0. It steepens toward the cluster center down to -1.4. This is the first time the spectral steepening is resolved across the southern relic. However, the steepening in the relic RS-1 is less clear. Most of RS-1 has a flat spectral index of -0.9. The relic RN also shows spectral steepening, but with a gentle gradient. The spectral index decreases from -0.9 at the northern edge to -1.1 at the center.

We also computed a spectral index map between 325 MHz and 3.0 GHz, where the 3.0 GHz image was published by \citet{2017ApJ...841....7B}. This map exhibits a similar spectral steepening trend as the spectral map between 325 MHz and 1.5 GHz, but with a lower spatial resolution. 

\begin{figure*} 
\begin{center}
\begin{tabular}{cc}
\includegraphics[height=2.5 in]{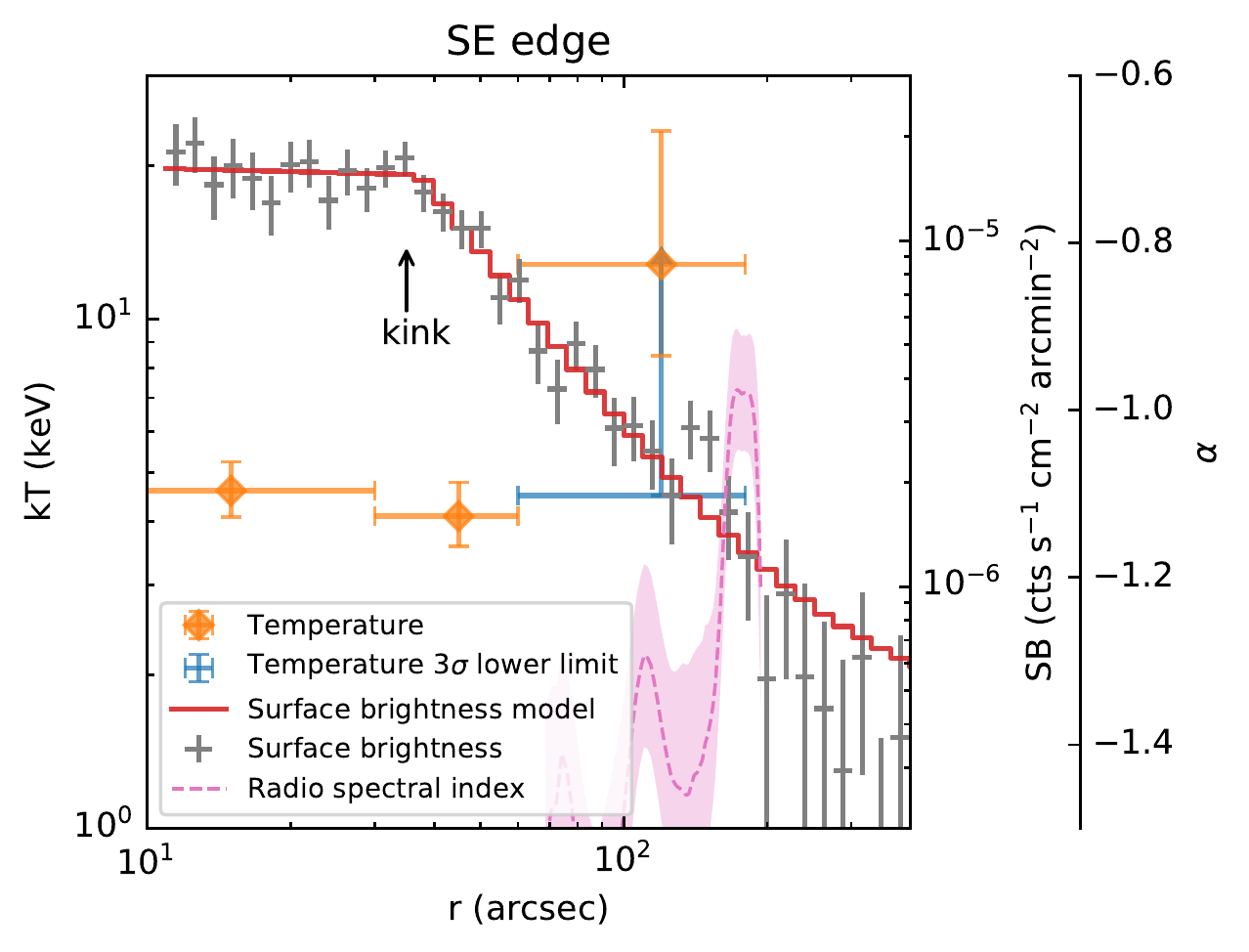} &
\includegraphics[height=2.5 in]{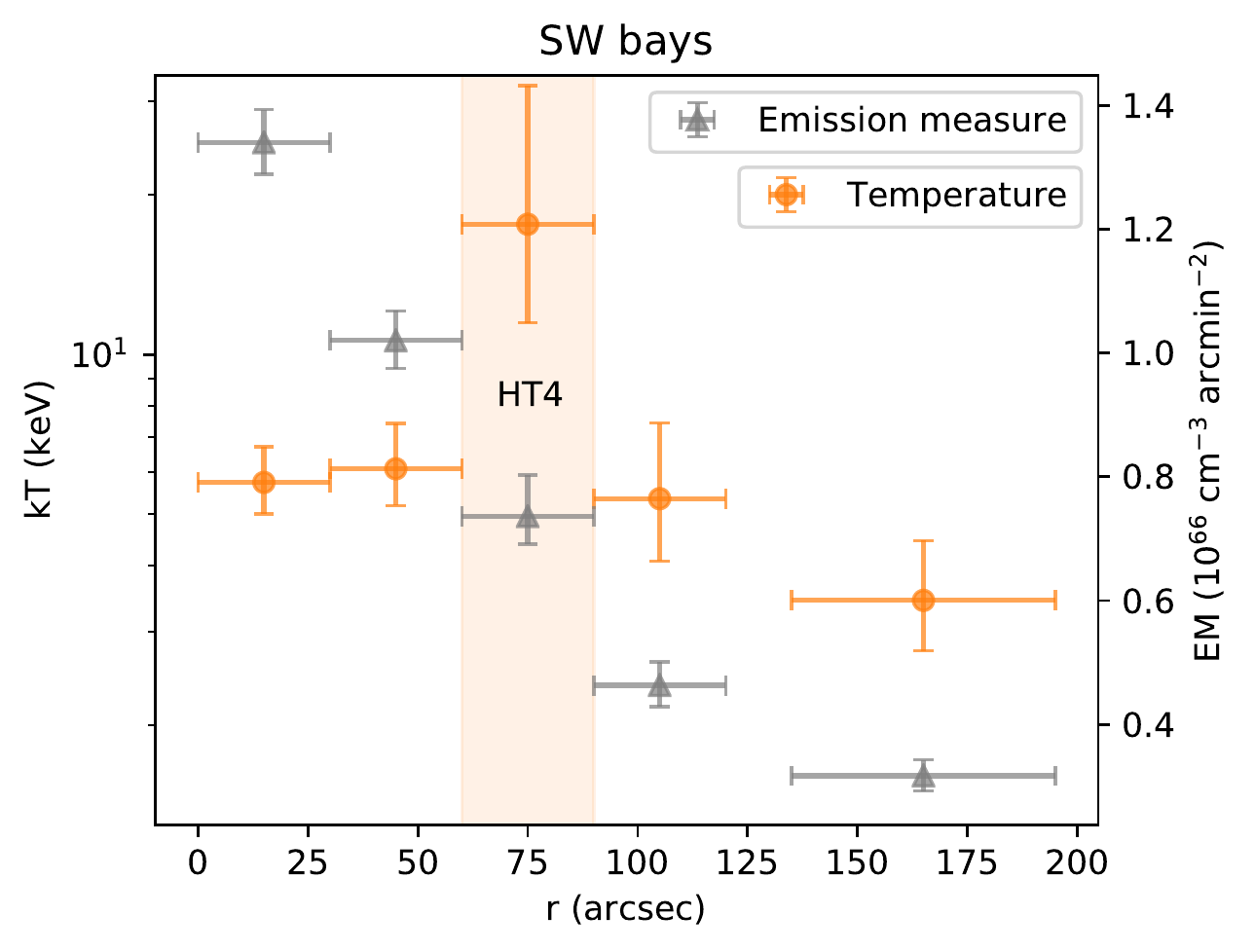}\\
\includegraphics[height=2.5 in]{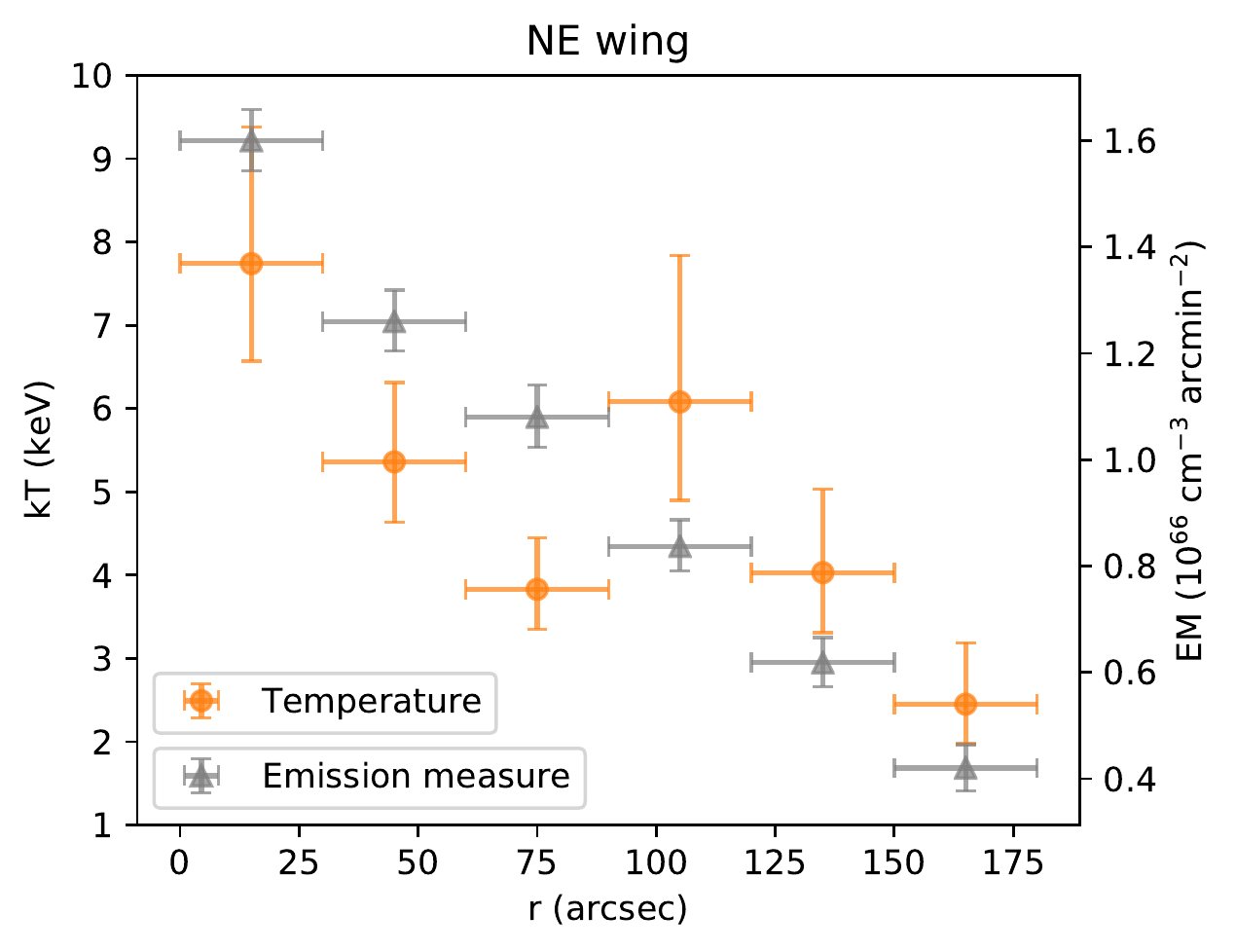} &
\includegraphics[height=2.5 in]{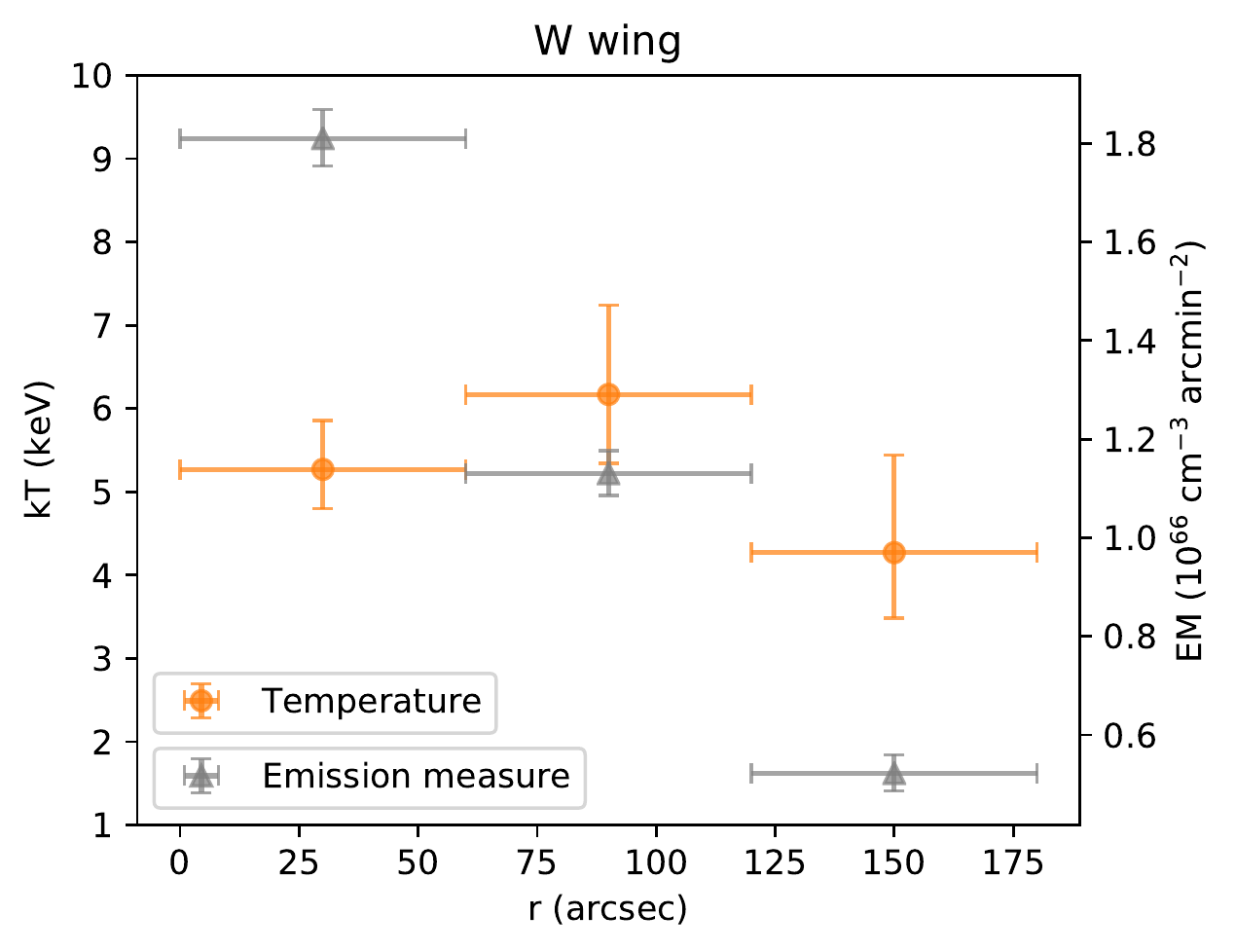}
\end{tabular}
\caption{
Temperature and surface brightness (or emission measure) profiles in the SE edge, SW bays, NE wing, and western wing extraction regions. The region definitions are shown in the bottom panel of Fig. \ref{fig:image}.
}
\label{fig:pro}
\end{center}
\end{figure*}

\section{Results}\label{sect:result}

\subsection{General X-ray properties}\label{sec:result-general}
We fit a global spectrum in a circular region with radius $r_{500}$, derived from the \ac{SZ} mass ($M_\mathrm{500,SZ}=5.2\pm0.4$ $M_\sun$, \citealt{2016A&A...594A..27P}), and centered at [23:43:42.1,
+00:17:49.9], which is close to the pressure peak and the X-ray centroid (see the dashed circle in Fig. \ref{fig:offset}). We obtained $kT=5.32\pm0.18$ keV and $Z=0.29\pm0.07$ $Z_\sun$, which are consistent with the results of previous Suzaku observation \citep{2013PASJ...65...16A}.


The previous shallow observation already shows that the X-ray emission is NW-SE elongated with substructures of an outbound bullet-like structure at the NW, a tail-like structure at the SE, and an excess emission toward the NE \citep{2014MNRAS.443.2463O}. The deep \emph{Chandra} observations reveal new surface brightness features, which are marked in the top right panel of Fig. \ref{fig:image}. The NE subcluster is actually a cone-shaped structure, where linear surface brightness edges are on the both sides of the cone. Below the cone, two wings stretch out toward the NE and SW directions, respectively. The apex of the cone is connected to the NE wing by a slim X-ray bridge. A linear surface brightness edge is in the SE direction, which was believed to be associated with the southern radio relic. In the SW direction, different from the SE linear structure, two surface brightness deficit regions are repeatedly shown.


Thermodynamic maps are plotted in Fig. \ref{fig:thermomap}. We also provide the temperature uncertainty map as in Fig. \ref{fig:terrmap}. Limited by total counts and the size of the object, the spatial bins are coarse. Nevertheless, the thermodynamic maps still imply possible merging features, for example, shocks, cold fronts, and core remnants, for which we customize the surface brightness and spectral extraction regions for detailed investigation in the following subsections.  In the temperature map, there are four regions where the temperatures are higher than the ambient regions, which are labeled as HT1--4. The region HT1 is close to the geometric center of the merging system, while HT1--3 are in relatively low density regions. HT2 is between the NW comet and NE wing, HT3 is outside the SE linear surface discontinuity, and HT4 is coincident with the two bays. In the pseudo pressure map, the pressure smoothly decreases from the HT1 to outer regions. There is no significant pressure jump across the inner edge of the HT2 and HT3 regions. There might be a sharp pressure gradient at the outer edge of the HT4 region and need to be further inspected. The entropy map shows three low entropy area, which correspond to the NW outbound cone (LK1), NE wing (LK2), and part of the SE linear structure (LK3). 
Because convection within clusters transfers low entropy gas to the bottom of gravitational potential wells and high entropy gas to the outskirts \citep[e.g.,][]{2002ApJ...576..601V}, low entropy gas is found in the cores of relaxed clusters.
Therefore, the low entropy in those regions implies that the origin of the gas is related to cores of subclusters before the merger. The NW cone is a subcluster core being stripped. The SE linear structure is the remnant of a totally disrupted core, which is associated with the subcluster moving toward the SE. 

\subsection{SE edge}
We extracted a 0.5--2.0 keV surface brightness profile across the SE linear structure using the region shown in the bottom panel of Fig. \ref{fig:image}. The surface brightness profiles are plotted in the top left panel of Fig. \ref{fig:pro}. The surface brightness profile is similar to the profile in Fig. 3 of \citet{2014MNRAS.443.2463O}, but with a factor of 2.4 higher signal to noise ratio. 
The deep observations reveal that the SE edge does not resemble other textbook examples of surface brightness discontinuities. 
The surface brightness profile is flat before a kink at $\sim40\arcsec$. After the kink, the surface brightness continuously decreases in a power law until a bump between the radii of $120\arcsec$ and $160\arcsec$. Although there is a significant change of the slope, there is no clear associated jump of density (see \citealt{2007PhR...443....1M} for examples with sharp density jumps). 
The profile can be well fitted by a unprojected broken power law model with $\mathrm{C}_\mathrm{stat}/\mathrm{d.o.f}=35.4/38$. The best-fit model was plotted as well. While using a single power law model, the fitting statistics is $\mathrm{C}_\mathrm{stat}/\mathrm{d.o.f}=179.2/40$, which is much worse than using a broken power law, indicating an excess of X-ray emission contributed by a gas clump above the underlying density profile.

Previous work on short observations was unable to explore temperature structures across the edge. With the deep observations, 
we tentatively search for potential temperature jumps.
We plotted the temperature profile in the same panel as the surface brightness profile in Fig. \ref{fig:pro}.
The temperature in the $0\arcsec$--$30\arcsec$ and $30\arcsec$--$60\arcsec$ bins are $4.6\pm0.5$ keV and $4.1\pm0.5$ keV, respectively, which are slightly below the averaged $kT_{500}=5.32\pm0.18$ keV. After that, it reaches $12.8^{+10.6}_{-4.8}$ keV in the $60\arcsec$--$180\arcsec$ bin. 
Outside this bin, the temperature cannot be constrained due to the fast decreasing surface brightness.
Based on the temperature profile, there is no evidence for shock heating on the inner side of the kink, which 
disfavors the previous speculation by \citet{2014MNRAS.443.2463O} that the edge is a shock front. 
The existence of an additional gas clump as well as its relatively low temperature suggest that it is the remnant of the core of the southern subcluster, which agrees with the low entropy in the region LK3 (see Fig. \ref{fig:thermomap}). The high temperature outside the cool clump is likely to be heated by the leading shock. The pattern of a hot post shock region followed by a cool gas clump has been found in many post merger systems, where the cool component could be a remnant core either remains compact (e.g., Bullet Cluster, \citealt{2002ApJ...567L..27M,2006ESASP.604..723M}) or has been broken up (e.g., Abell 520, \citealt{2016ApJ...833...99W}). In our case, the core remnant is more diffuse, implying that either the progenitor is not a strong cool core or the diffusion processes has already taken place.
We also estimated the $3\sigma$ lower limit of the temperature in $60\arcsec$--$180\arcsec$ bin by using the $\Delta\mathrm{C_{stat}}=9$ criterion. The $3\sigma$ lower limit is $4.5$ keV. We use this lower limit to constrain the Mach number of the undetected shock that is possible accelerating particles at the outer edge of the southern relic in Sect. \ref{sect:shock}.

The location of the shock front is unable to be resolved in X-rays due to the limited photons. Nevertheless, the functions of shocks in terms of particle acceleration shed light on the possible location. We plot the spectral index of the radio relic in the same panel. The radio relic has a relative flat spectrum at the outer edge ($r\sim200\arcsec$) and the spectral index decreases toward the cluster center. Based on the shock-relic connection established for most radio relics \citep[e.g.,][]{2010ApJ...715.1143F}, the position of the shock coincides with the flat edge, which is about $160\arcsec$ (640 kpc) away from the cool gas clump. 




\subsection{SW bays}
The locations of the two bays are coincident with a high temperature region in the temperature map (HT4). The gas outside this region has a lower temperature (<4 keV). To locate the hot region in the NE-SW direction and search for possible temperature jumps, we extracted temperature profiles from the merging axis to the SW outskirts, the region configuration is defined in the bottom panel of Fig. \ref{fig:image}. In addition, we extracted an emission measure profile that presents the trend of surface brightness (see the top right panel of Fig. \ref{fig:pro}). The third bin, ranging from $60\arcsec$ to $90\arcsec$ and covering most of the area of the two bays, has an extreme high temperature of $17.6^{+14.5}_{-6.1}$ keV, which is a factor of $2.9^{+2.4}_{-1.2}$ higher than the second bin. After this bin, the temperature decreases to the average value of the cluster and reaches $3.4\pm0.8$ keV in the outermost bin.

\begin{figure}[]
\begin{center}
\begin{tabular}{l}
\includegraphics[width=0.85\hsize]{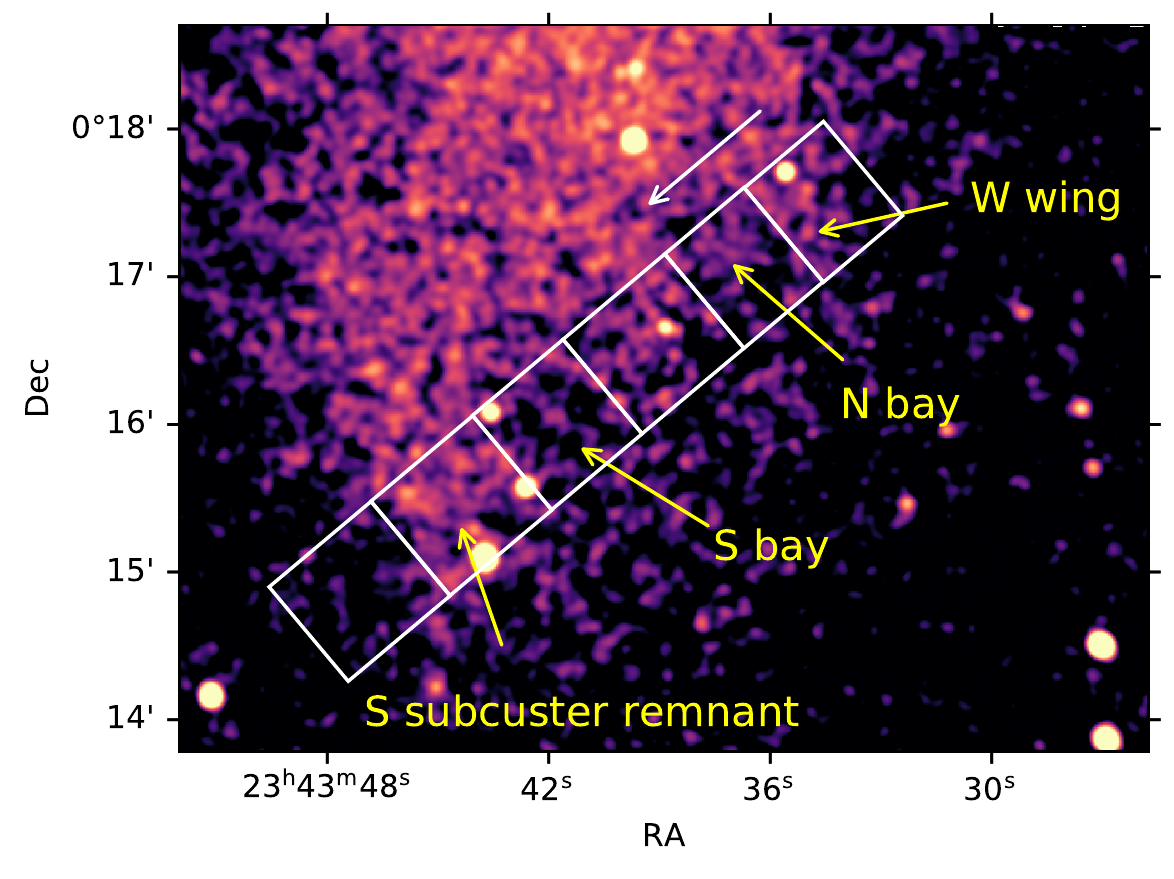}\\
\includegraphics[width=0.95\hsize]{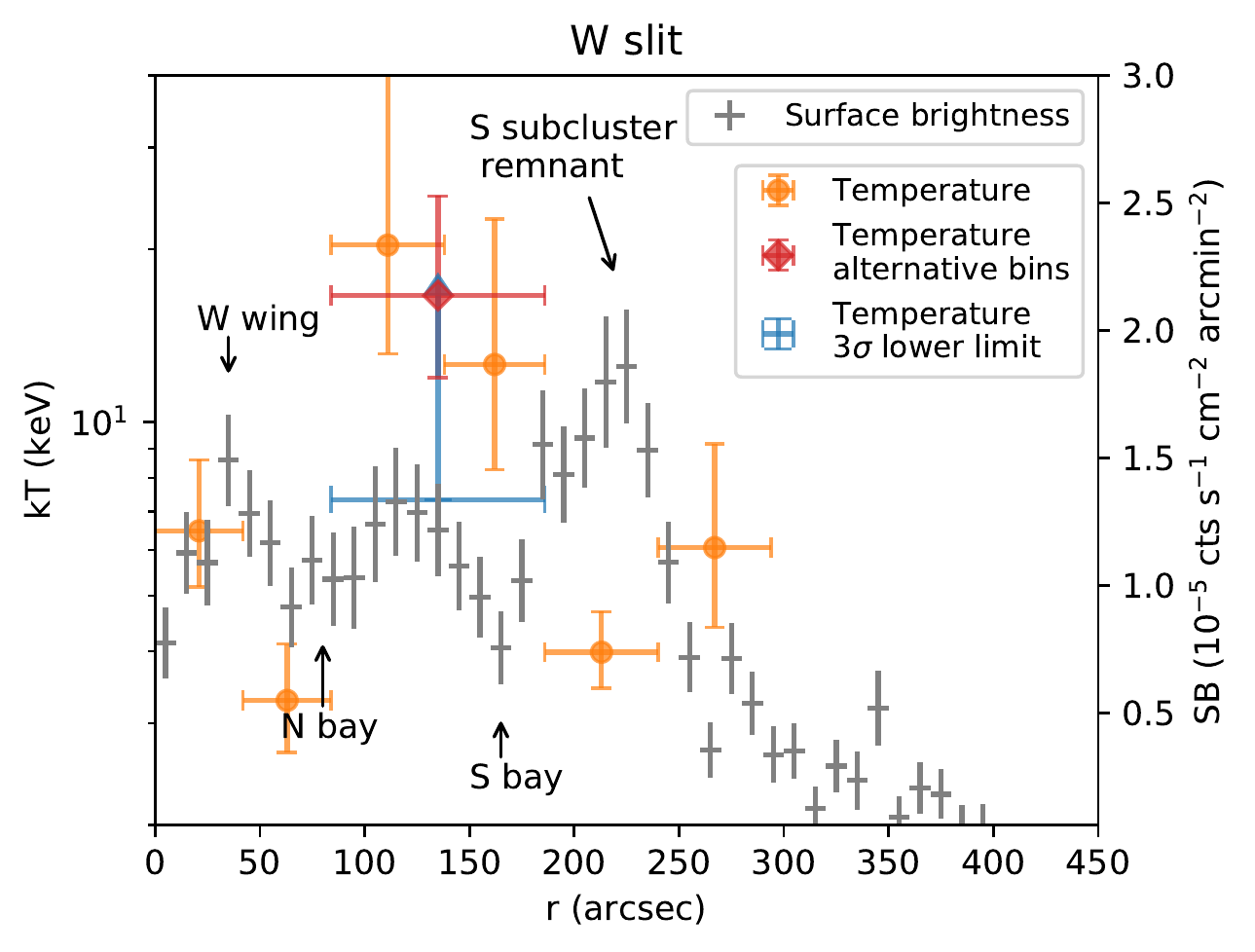}
\end{tabular}
\caption{
\emph{Top:} Zoomed-in view of the western flank of the cluster. Individual temperature bins are plotted in white. The arrow indicates the direction of profiles.
\emph{Bottom:} Temperature and surface brightness profiles extracted from the western slit region.
}
\label{fig:w_slit}
\end{center}
\end{figure}

The third temperature bin in the top right panel of Fig. \ref{fig:pro} includes both bays. The profile in the NE-SW direction is unable to probe the origin of the hot gas and distinguish the thermodynamic properties of the individual bays. Therefore, we define perpendicular extraction regions along the NW-SE direction (see the top panel of Fig. \ref{fig:w_slit}). It starts from the western wing and goes through two bays and the remnant of the southern subcluster. The surface brightness and temperature profiles in the western slit region are shown in the bottom panel of Fig. \ref{fig:w_slit}. The temperature profile presents entirely different thermodynamic properties of the two bays. While the \ac{ICM} in the northern bay is $3.3\pm0.7$ keV, the southern bay temperature is $12.6^{+10.0}_{-4.3}$ keV. Furthermore, the ridge between the two bays has the highest temperature in the profile. Considering that the ridge has higher surface brightness than the two bays, which indicates a higher density, together with the high temperature, the gas in the two bays and the ridge does not hold pressure equilibrium. We measure the temperature by combining the two hot bins and obtain a $16.6^{+8.1}_{-4.7}$ keV temperature. Similar to the hot bin in the SE post shock region, we put a $3\sigma$ lower limit of $7.3$ keV for this bin, which is still a factor of two higher than the temperature in the northern bay.

\subsection{NE and western wings}
The temperature and emission measure profiles of the NE and western wing regions are plotted in the bottom left and bottom right panels of Fig. \ref{fig:pro}, respectively. The spectral extraction regions are designed to ensure at most $20\%$ uncertainty for each temperature measurement. Therefore, we have six bins for the NE wing but only three bins for the western wing.

By visual inspection, we find no significant surface brightness jump across the NE wing, which is supported by the monotonously decreasing emission measure profile.
However, the temperature profile exhibits features that are not seen in the surface brightness. The temperature in the innermost bin is $7.7\pm1.2$ keV, which is consistent with the HT1 region in the temperature map. Until $r=90\arcsec$, the temperature drops monotonously, after which there is a sudden rise from $3.8\pm0.5$ keV to $6.1\pm1.2$ keV, where the jump has a significance of $1.8\sigma$. This plausible temperature enhancement does not correspond to any surface brightness jump. After the jump, the temperature continuously decreases to $2.5\pm0.5$ keV at the wingtip. 

On the contrary, the western wing does not show strong temperature variation, though the spatial bins are coarse due to the small size. The temperature at the wingtip is $4.7\pm0.9$ keV, which is closer to the averaged temperature of the cluster than the NE wingtip.

\subsection{NW cone}

\begin{table*}
\caption{Best-fit parameters of the surface brightness profiles across the cone.}
\begin{center}
\begin{tabular}{cccccccc}
\hline\hline
 & $n_\mathrm{e}$ & $\alpha_1$ & $\alpha_2$ & $\phi_\mathrm{b}$ & $C$ & $\theta_\mathrm{axis}$ & C-stat / d.o.f\\
  & $10^{-3} $cm$^{-3}$ &  &  & $\degr$ & & $\degr$ & \\
 \hline
 $100\degr$--$200\degr$ & $1.71\pm0.14$ & $-0.22\pm0.15$ & $0.92\pm0.13$ & $19.9^{+0.7}_{-1.3}$ & $1.56\pm0.23$ & $135.6\pm0.6$ & $54.2/44$\\
 $100\degr$--$136\degr$ & $1.60^{+0.06}_{-0.10}$ & 0 (fixed) & $1.70\pm0.4$ & $20.5^{+3.2}_{-0.6}$ & $1.12^{+0.27}_{-0.12}$ & 136 (fixed) & $12.1/14$\\
  $136\degr$--$200\degr$ & $1.46\pm0.08$ & 0 (fixed) & $0.89\pm0.12$ & $19.5\pm1.7$ & $1.26^{+0.07}_{-0.26}$ & 136 (fixed) & 27.9/28\\
 \hline
\end{tabular}
\end{center}
\label{tab:cone}
\end{table*}

\begin{figure}
\begin{center}
\includegraphics[width=0.99\hsize]{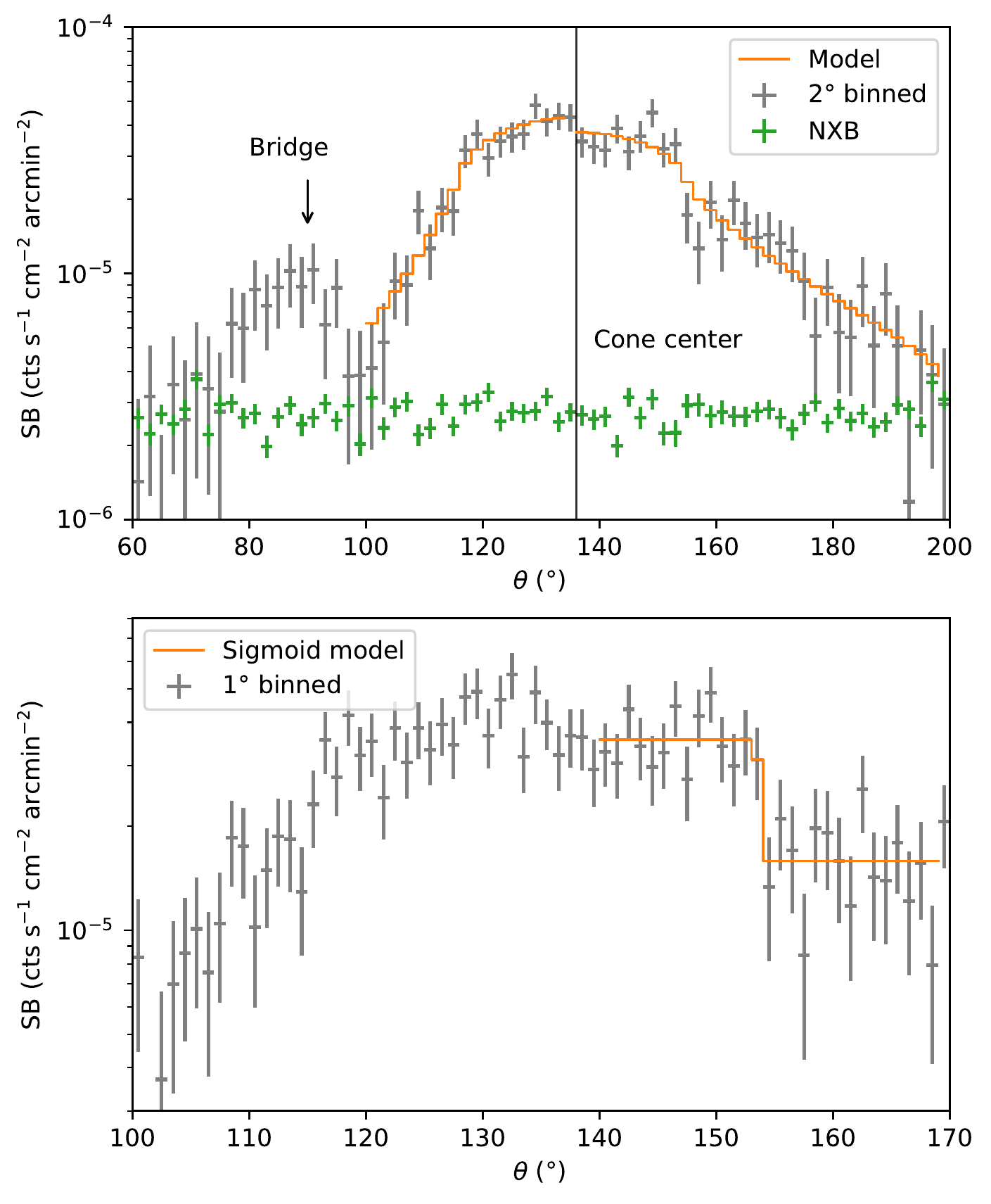}
\caption{
\emph{Top:} Azimuthal surface brightness profile with a bin width of $2\degr$. The vertical line is the axis of the cone. The $100\degr$--$136\degr$ and $136\degr$--$200\degr$ profiles are fitted separately.
\emph{Bottom:} Zoomed-in view of the profile with a bin width of $1\degr$ together with the best-fit sigmoid model from $140\degr$--$170\degr$. 
}
\label{fig:cone}
\end{center}
\end{figure}

The low entropy of the gas in the cone, the higher temperature of the ambient gas, and the equilibrium of the pressure across the cone show that this is an outbound subcluster core. Unlike any other observed remnant core cold front, this remnant core is not in a blunt-body shape, but in a cone shape. On both side of the cone, the cold fronts are linear and with projected lengths of $\sim400$ kpc. 
Unlike typical radial profiles, we extract a radially averaged surface brightness profile using sectors of annulus centered on the apex of the cone. The inner and outer radius are $50\arcsec$ and $100\arcsec$, respectively (see the bottom panel of Fig. \ref{fig:image}). We define the azimuthal angle starting from the north and increasing eastward. The $2\degr$ binned profile is plotted in the top panel of Fig. \ref{fig:cone}. The bump centered at $\sim85\degr$ is the ``bridge'' connecting the cone and the eastern wing. The surface brightness continuously increases from $100\degr$ to $115\degr$. From $115\degr$, which is the left edge of the cone, the surface brightness is relatively flat until a sudden drop at $\sim150\degr$. After the drop, the surface brightness is in a plateau for $\sim15\degr$ then gradually fades away. 

To probe the gas density inside the cone, we tentatively fit the profile using a projected broken power law model. The detailed description of the model and the projection of the conic body is in Appendix \ref{app:cdpl}. 
We first fit the entire profile from $100\degr$ to $200\degr$ to measure the width of the cone and the position angle of the axis. The center of the cone is at $\theta_\mathrm{axis}=135\degr.6\pm0.6$ and the gas density has a break at $\phi_\mathrm{b}=19\degr.9^{+0.7}_{-1.3}$. However, a single component model cannot fit both sides of the conic subcluster well. The left side has a steeper gradient than the right side. 
Meanwhile, the best-fit power law index inside the cone $\alpha_1$ is negative, which indicates that the gas density at the edge is higher than the center. 
Therefore, we fit the $100\degr$--$136\degr$ and $136\degr$--$200\degr$ profiles separately. The parameter $\alpha_1$ is fixed to zero to investigate how much the observed profile deviates from a uniform gas distribution. The best-fit parameters of each profile are listed in Table \ref{tab:cone}. 
The left part can be well fitted with a uniform gas density in the cone. 
On the right part, the model cannot fit the profile at the jump well. The surface brightness enhancement before the jump indicates a high density shell at the cone edge. 

We note that the surface brightness discontinuity on the right side of the axis is sharp. To further quantify the location of the jump, we binned the profile with a width of $1\degr$ (see the bottom panel of Fig. \ref{fig:cone}). The jump is at $\sim154\degr$ in the $1\degr$ binned profile. On both sides of the jump, the profiles are flat. We tentatively fit the profile in the $140\degr$ -- $170\degr$ range to determine the jump location as well as the jump width using a sigmoid model.
We adopted the form of error function, which is naturally a Gaussian convolved box function. In our case, the model is 
\begin{equation}
    S(\theta) = S_0 - \frac{S_{\mathrm{jump}}}{2}\times\Phi\left(\frac{\theta-\theta_{\mathrm{jump}}}{\sqrt{2}\sigma}\right),
\end{equation}
where $S_0$ and $S_{\mathrm{jump}}$ are profile normalization and jump strength, $\Phi{x}$ the error function, $\theta_{\mathrm{jump}}$ the jump location, and $\sigma$ the width of the Gaussian smooth kernel. The best-fit jump location is $154\degr.2\pm0.8$ and the best-fit kernel width is $0\degr.07^{+2.28}_{-0.07}$. 

\subsection{Bridge and northern outskirts}
\begin{figure}[!]
\begin{center}
\includegraphics[width=0.99\hsize]{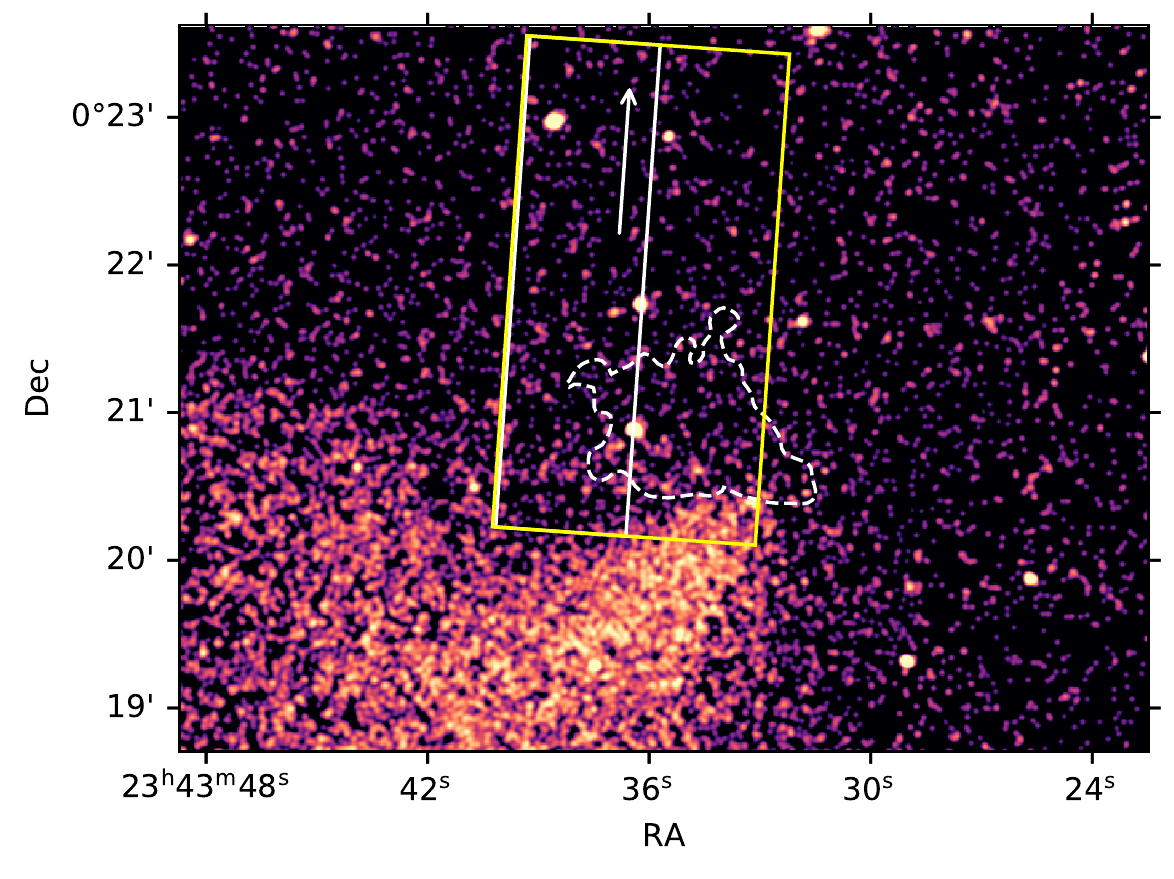}\\
\includegraphics[width=0.99\hsize]{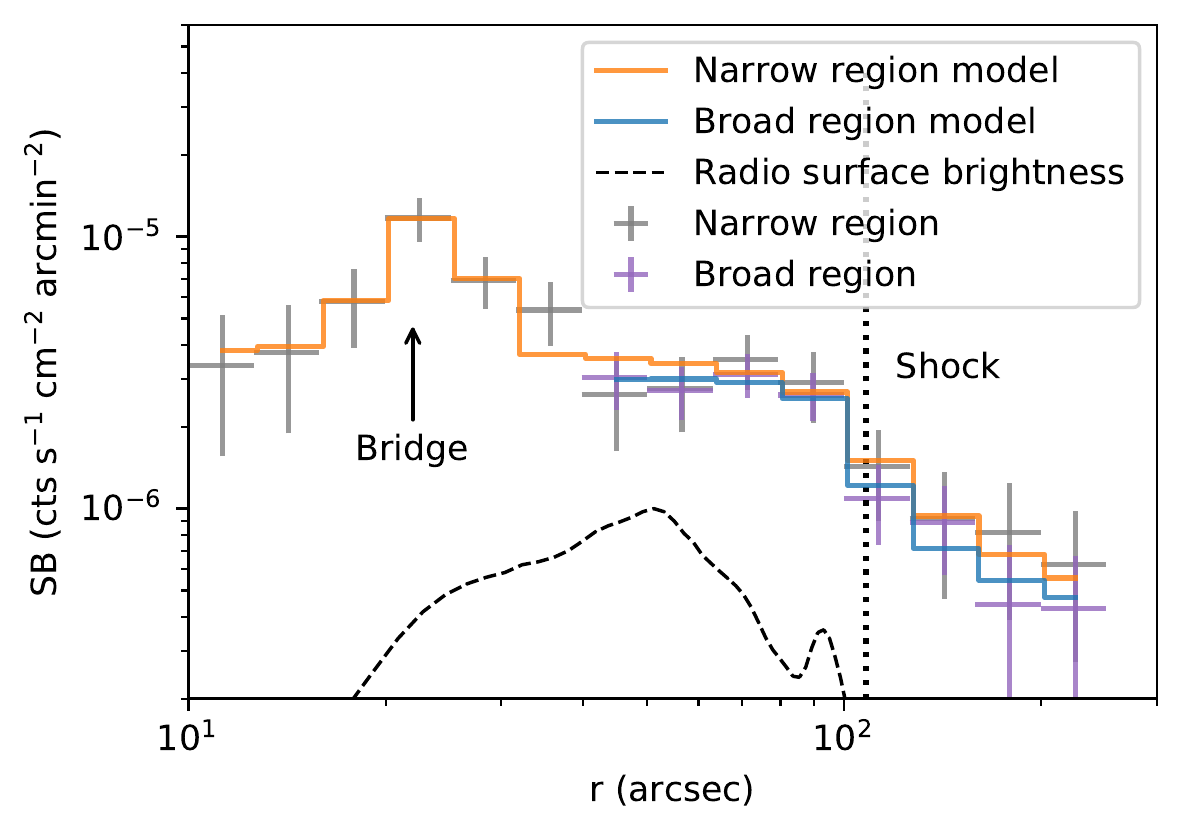}
\caption{
\emph{Top:} Narrow (white) and broad (yellow) extraction regions for surface brightness profiles in the northern outskirts. The dashed contour is the northern relic in the GMRT 325 MHz image at the $3\sigma_\mathrm{rms}$ level.
\emph{Bottom:} Surface brightness profiles and their best-fit model in the narrow and broad regions. The bridge and the surface brightness jump are located at $r=23\arcsec$ and $r=108\arcsec$, respectively. The GMRT 325 MHz profile is plotted with a dashed curve, the outer edge of which is associated with the potential X-ray jump.
}
\label{fig:bridge}
\end{center}
\end{figure}

We extracted the surface brightness profile across the bridge from the white narrow box region (the left half of the yellow broad region) in the top panel of Fig. \ref{fig:bridge}. The surface brightness profile is plotted in the bottom panel of Fig. \ref{fig:bridge}.
The bridge is located at $r\sim20\arcsec$, after which the surface brightness keeps flat until a plausible jump at $r\sim100\arcsec$. To check the significance of the bridge and the jump, we fit the profile using a compound model that consists of a Gaussian profile and a projected double power law density profile. The best-fit surface brightness of the Gaussian component is $(9.1\pm3.0)\times10^{-6}$ cts s$^{-1}$ cm$^{-2}$ arcmin$^{-2}$, which means that the gas bridge has a $3\sigma$ detection. The best-fit density jump is $1.6\pm0.6$, which is unconstrained due to the limit of counts. However, the best-fit power law index changes from $0.0\pm0.1$ before the jump to $0.8\substack{+1.0\\-0.1}$ after the jump. We extracted a surface brightness profile using a broad region (the yellow box in the top panel of Fig. \ref{fig:bridge}) to check whether the potential jump is associated with the northern relic. The surface brightness in the broad region is plotted in the bottom panel of Fig. \ref{fig:bridge} as well. The profile has a similar trend as that in the narrow region and the best-fit jump location is $108\arcsec\pm4$, which is close to the outer edge of the northern relic. The best-fit density jump of the broad region profile is $1.8\pm1.1$, which means that the density jump still cannot be constrained.

\section{Discussion}\label{sect:discussion}
\subsection{Shock fronts and radio relics}\label{sect:shock}
By measuring the temperature profile, we discovered that the previous SE surface brightness jump reported by \citet{2014MNRAS.443.2463O} is not a shock but rather a kink contributed by cool gas from the remnant core. Outside the kink, the X-ray surface brightness drops dramatically, preventing us from investigating the exact location of the shock in the X-rays. Nevertheless, at radio frequencies, the 325 MHz - 1.5 GHz spectral index map clearly shows the spectral steepening from the edge toward the cluster center. It is the first time that we see the evidence of shock acceleration and subsequent radiative cooling of the relativistic plasma in the radio relic of this cluster. The spectrally flat edge of the southern relic, which is $\sim160\arcsec$ away from the kink, is the possible location of the southern shock. 

In the northern relic, the spectral index is flat at the northern edge, where the X-ray surface brightness changes its gradient. Although the density jump cannot be constrained, the variation in the gradient at this position suggests that the northern shock that is responsible for (re)accelerating the northern relic is at this position.

Based on the \ac{DSA} theory, the radio injection spectral index is a function of the Mach number, 
\begin{equation}\label{eq:radio_mach}
\alpha_\mathrm{inj}=\frac{\mathcal{M}^2+3}{2\left(\mathcal{M}^2-1\right)}.
\end{equation}
Since the spectral index is spatially resolved, this relation can be directly applied to the southern relic, where the injection spectral index is $1.00\pm0.06$. We obtain a radio Mach number $\mathcal{M}_\mathrm{radio,S}=2.2\pm0.1$. This number is lower than using the integrated spectral index $\alpha_\mathrm{S,int}=-1.20\pm0.18$ \citep{2017ApJ...841....7B} and assuming the relation between the injection and integrated spectral indices is $\alpha_\mathrm{int}=\alpha_\mathrm{inj}-0.5$ \citep{1962SvA.....6..317K}. A theoretical investigation shows that this simple assumption does not hold when the shock is nonstationary, which is the case of cluster merger shocks \citep{2015JKAS...48....9K}. We also applied Eq. \ref{eq:radio_mach} to the northern relic. The updated radio Mach number $\mathcal{M}_\mathrm{radio,N}=2.4\pm0.4$.

Meanwhile, the Mach number can be estimated by using the temperature jump based on the Rankine-Hugoniot condition \citep{1959flme.book.....L},
\begin{equation}\label{eq:t_mach}
\frac{kT_\mathrm{post}}{kT_\mathrm{pre}}=\frac{5\mathcal{M}^4+14\mathcal{M}^2-3}{16\mathcal{M}^2}.
\end{equation}
The pre-shock region is too faint to measure the temperature. Therefore, we tentatively extrapolate it by using a universal temperature profile \citep{2010ApJ...721.1105B}, which remarkably agrees with \emph{Suzaku} observations of galaxy cluster outskirts. The profile is
\begin{equation}
\frac{kT(r)}{kT_\mathrm{ave}}=(1.74\pm0.03)\times\left[1+(0.64\pm0.10)\times\left(\frac{r}{r_{200}}\right)\right]^{-3.2\pm0.4}.
\end{equation}
The spectrally flat boundary of the relic is located at $r\sim0.7r_{200}$, at which the predicted pre-shock temperature is $2.8\pm0.6$ keV. Because the temperature measurement in the post shock region has a large error bar, we only put a lower limit for that Mach number. By combining the $3\sigma$ lower limit of the post-shock temperature $4.5$ keV, we obtain the $3\sigma$ lower limit of the X-ray Mach number 1.6. The estimated lower limit of the X-ray Mach number matches the value of the radio Mach number. Because we can only put a lower limit of the X-ray Mach number, it is not clear whether this shock is another case with Mach number discrepancy between the X-ray and the radio. Unfortunately, in the X-ray, the temperature between the bridge and the potential jump in the northern outskirts (see Fig. \ref{fig:bridge}) is not constrained. 

\subsection{Conic subcluster}
The shapes of the subclusters in other merging systems usually show a spheroid core, and the surface brightness decreases radially from the center of the core. As an exception, the subcluster in ZwCl 2341+0000 does not have a concentrated surface brightness peak and the radius of curvature at the apex can be limited to be less than 30 kpc. As a comparison, the straight cold fronts on both sides have lengths of 400 kpc. Meanwhile, by using a surface brightness profile fit, we estimated that the gas density at the center of the cone is about $10^{-3}$ cm$^{-3}$, which is relatively low among all cool core and non cool core clusters \citep[e.g.,][]{2010A&A...513A..37H}. The corresponding cooling time is about 27 Gyr, which is much larger than the Hubble time, implying an expansion of the core after the collision with another subcluster.

An analogy of the cone-shaped subcluster in simulations is presented in Fig. 9 of \citet{2019MNRAS.482...20Z} at $t=0.03$ Gyr. As a bullet-like subcluster leaves the atmosphere of the main cluster, due to the decrease in pressure, the far end starts to expand while the two sides are still under ram pressure stripping. Thus, the bullet-like subcluster is elongated and deformed. A cone-shaped subcluster with a narrow aperture appears at a certain stage of the merging process. We note that in ZwCl 2341+0000, the projected brightest cluster galaxy position is outside but very close to the apex of the cone, indicating that the dark matter sub-halo is still leading the gaseous subcluster. As the merger goes on, the tip of the cone will be continuously expanding to form a slingshot cold front \citep[e.g., Abell 168][]{2004ApJ...610L..81H}. Another similar analogy is the $R=1:3$ and $b=0$ kpc simulation set in \citet{2011ApJ...728...54Z}\footnote{\url{http://gcmc.hub.yt/fiducial/1to3_b0/index.html}}. At $t=1.6$ Gyr, the subcluster evolves to be a cone with an aperture of $40\degr$, which is similar with the observed conic structure. The fate of the conic subcluster in this simulation is the same as the one in \citet{2019MNRAS.482...20Z}. After this epoch, the tip of the cone starts to expand in the low density outskirts. This simulation also produces a dense gas shell at the cone boundary, which is similar to the right part of the cone in ZwCl 2341+0000. 

\begin{figure}[]
\begin{center}
\includegraphics[width=0.99\hsize]{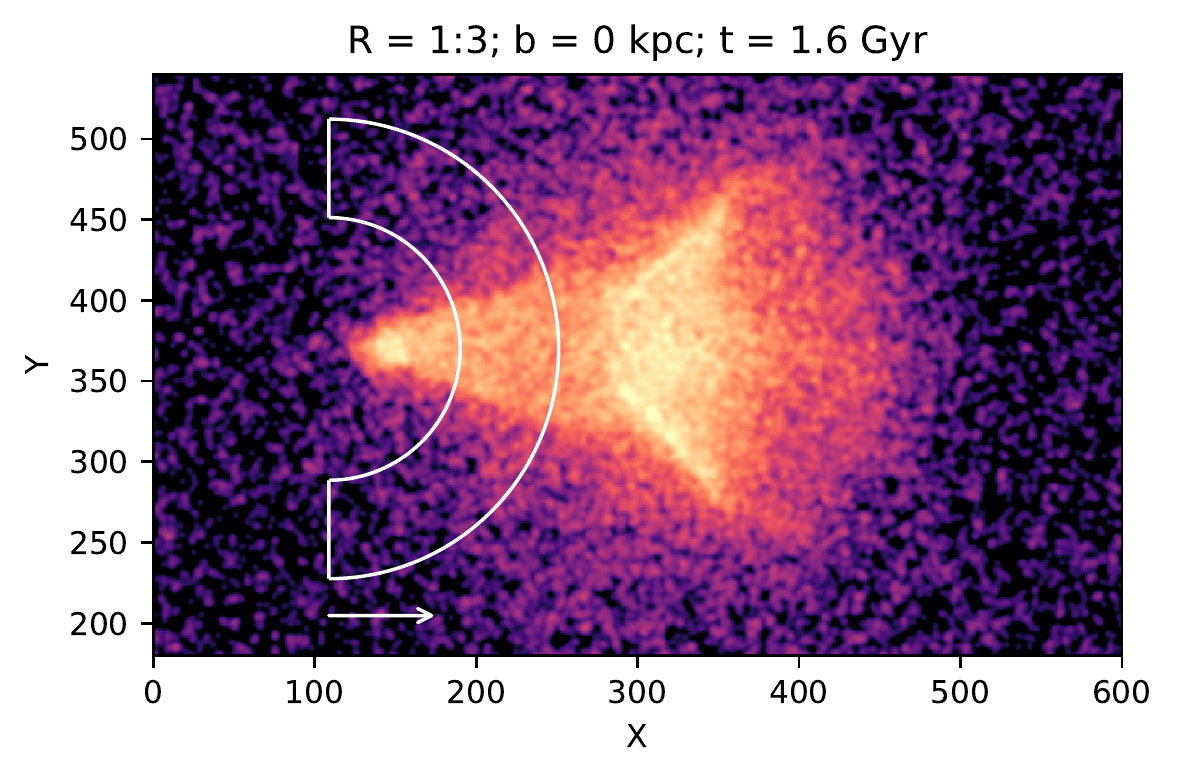}\\
\includegraphics[width=0.99\hsize]{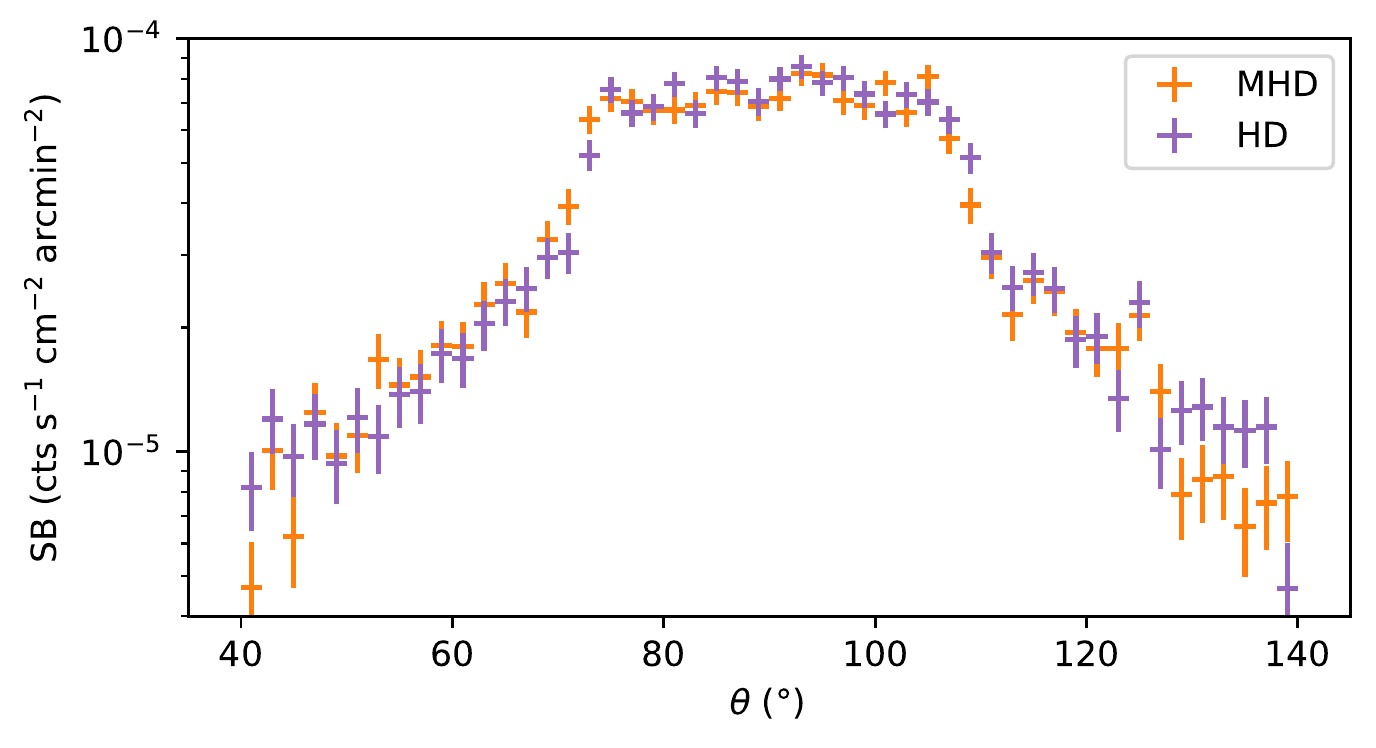}
\caption{
\emph{Top:} Mock X-ray image of a cone-shaped subcluster in the $\beta=100$ MHD simulation, whose merger configuration is the same as in \citet{2011ApJ...728...54Z} and \citet{2019ApJ...883..118B}. In this snapshot, two wing-like structures in addition to the cone are reproduced as well.
\emph{Bottom:} Azimuthal surface brightness profiles extracted from the mock observations of the MHD (orange) and HD (purple) simulations.}
\label{fig:simulation}
\end{center}
\end{figure}


\subsubsection{Effect of magnetic fields}
During a merger, the magnetic field in the cluster can be amplified through multiple processes and could produce observable features such as plasma depletion layers \citep[see][for a review]{2018SSRv..214..122D}.
We investigate the effect of magnetic field in forming such conic subclusters by comparing a pure \ac{HD} simulation and a \ac{MHD} simulation.
With the same merger configurations as \citet{2011ApJ...728...54Z}, \citet{2019ApJ...883..118B} carried out \ac{MHD} simulations with initial condition $\beta=200$ \footnote{$\beta\equiv p_\mathrm{th}/p_\mathrm{B}$}. Similarly, we ran a new \ac{MHD} simulation of the $1:3$ mass ratio with $\beta=100$, where the effect of magnetic field should be more significant than the $\beta=200$ simulation. We created mock images by putting both the \ac{HD} and MHD simulations at $z=0.27$  with ACIS configuration and 200 ks exposure time. The mock image of the MHD simulation is shown in the top panel of Fig. \ref{fig:simulation}. 

We extracted azimuthal surface brightness profiles for the simulated conic subclusters. The profiles are plotted in the bottom panel of Fig. \ref{fig:simulation}. The \ac{MHD} simulation has a similar profile as the pure \ac{HD} simulation, which means that the magnetic field with a $\beta=100$ initial condition does not play a significant role in shaping the conic subcluster. Moreover, we do not find significant plasma depletion layers that push gas out and exhibit dark streaks in the image (as was found by, e.g., \citealt{2016MNRAS.455..846W} in a deep observation of the Virgo sloshing cold front). Therefore, the observed conic structure is consistent with both HD and MHD simulations and the effect of magnetic fields cannot yet be ascertained with the current data.

\subsubsection{Gas bridge and \ac{ICM} viscosity}
In addition to the rare conic subcluster, the X-ray bridge structure is seldom discovered. The temperature of the bridge itself is $4.3\pm1.1$ keV and that of the combination of the bridge and the dark pocket below the bridge is $8.6\pm2.4$ keV. The similar temperature of the bridge and the cone indicates that the bridge could be the stripped gas trail from the cone. A similar structure has been found in \object{M 89}, where two horns are attached to the front of the remnant core \citep{2006ApJ...644..155M}. \citet{2017ApJ...848...27K} argues that the horns could either be \ac{KHI} eddies with a viewing angle of $30\degr$ or be due to the previous active galactic nucleus activities. Another similar case is galaxy group \object{UGC 12491} \citep{2012ApJ...754..147R}, where the stripped tails split from the core. 
Although our cluster has an order of magnitude larger physical scale than the cases of M 89 and UGC 12491, because HD is scale-free, the \ac{KHI} could still be responsible for the stripped gas trail in our system. If this gas trail is from a \ac{KHI} eddy, based on its 400 kpc length, it should be stripped at the time when the stagnation point of the subcluster was about 400 kpc closer to the system centroid, and then evolves together with the subcluster.

Viscosity suppresses the development of \ac{KHI} eddies that are smaller than a critical scale. In turn, with the existence of an eddy at a certain scale, we are able to constrain the upper limit of the viscosity $\mu$, whose expression is given by \citet{2021MNRAS.504.2800I}:
\begin{align}
    \mu\lesssim&\frac{\rho\lambda V}{a\sqrt{\Delta}}\nonumber\\
    \sim& 6300\text{ g cm}^{-1} \text{s}^{-1}\left(\frac{n_\mathrm{out}}{3\times10^{-3}\text{ cm}^{-3}}\right)\left(\frac{\lambda}{100\text{ kpc}}\right)\left(\frac{V}{1700\text{ km s}^{-1}}\right)\nonumber\\
    &\times\left(\frac{a}{10}\right)^{-1}\left(\frac{2.5}{\sqrt{\Delta}}\right),
\end{align}
where $n_\mathrm{out}$ is the ICM density outside the cold front, $\lambda$ the scale of the eddy, $V$ the shear velocity, $a$ a coefficient to be 10 for a conservative estimation, $\Delta\equiv(\rho_1+\rho_2)^2/(\rho_1\rho_2)$ calculated using the gas densities inside and outside the cold front. In our case, $n_\mathrm{out}\sim5\times10^{-4}$ cm$^{-3}$, which is about a factor of three lower than that in the cone. The scale of eddy is 400 kpc. We take the collision speed at pericenter $1900$ km s$^{-1}$ \citep{2017ApJ...841....7B} to be an upper limit of the shear velocity. We obtain an upper limit of viscosity of $\sim5000$ g cm$^{-1}$ s$^{-1}$. The full isotropic Spitzer viscosity is 
\begin{equation}
\mu_\mathrm{SP}=21000\times\left(\frac{kT}{14.7 \text{ keV}}\right)^{5/2} \text{ g cm}^{-1}\text{ s}^{-1},    
\end{equation}
assuming Coulomb logarithm $\ln\Lambda=40$ \citep{1956pfig.book.....S,1988xrec.book.....S}. A $\sim9$ keV ICM has $\mu_\mathrm{SP}\sim6200$ g cm$^{-1}$ s$^{-1}$, which means the viscosity is suppressed at least by a factor of 1.2 if the gas trail has a \ac{KHI} origin. With only this long gas trail, we cannot place strict constraints on the viscosity suppression factor, which is at least three in other systems, for example, three for \object{Abell 2319} \citep{2021MNRAS.504.2800I}, five for \object{Abell 2142} \citep{2018ApJ...868...45W}, and 20 for \object{Abell 3667} \citep{2017MNRAS.467.3662I}.


\subsubsection{Suppression of diffusion}
We note that the western side of the cone has a sharp surface brightness jump, whose width has $1\sigma$ upper limit of $2.4\degr$ azimuth.
Though this cluster has a relatively high redshift of 0.27, at which the projected linear scale corresponding to a $2.4\degr$ bin at the radius of $100\arcsec$ is 17 kpc, it is still worthwhile to inspect whether the diffusion is suppressed to form the observed sharp edge.
Assuming the \ac{ICM} is in thermal equilibrium, the mean free path of electrons is given by \citet{1988xrec.book.....S},
\begin{equation}
\lambda_e=31\ \mathrm{kpc}\left(\frac{kT}{10\mathrm{keV}}\right)^2\left(\frac{n_e}{10^{-3}\mathrm{cm}^{-3}}\right)^{-1}.
\end{equation}
The mean free path is about $4.1$ kpc for a $kT=4.5$ keV and $n_e=1.5\times10^{-3}$ cm$^{-3}$ \ac{ICM} inside the cold front, and is about $13.1$ kpc for a $kT=6.5$ keV and $n_e=1\times10^{-3}$ cm$^{-3}$ ICM outside the cold front.
Meanwhile, the mean free path of the particles moving from one side of the cold front to the other side is given by \citet{2001ApJ...551..160V},
\begin{equation}
\lambda_{\mathrm{A}\rightarrow\mathrm{B}}=
\lambda_\mathrm{A}\frac{kT_\mathrm{A}}{kT_\mathrm{B}}
\frac{G(1)}{G\left(\sqrt{kT_\mathrm{A}/kT_\mathrm{B}}\right)},
\end{equation}
where $G(x)=\left[\Phi(x)-x\Phi'(x)\right]/2x^2$ and $\Phi(x)$ is the error function, the subscripts A and B represent inside and outside of the cold front, and vice versa. The mean free path for particles moving from inside to the outside is $\lambda_{\mathrm{in}\rightarrow\mathrm{out}}=9.2$ kpc. The unsuppressed Coulomb diffusion will smear the density discontinuity by several mean free paths, which is comparable with the $1\sigma$ upper limit of the projected jump width. Therefore, without future deeper high spatial resolution observations, we cannot draw a solid conclusion whether the diffusion is suppressed.


\subsection{The origin of the NE wing}
In some head-on mergers, gas plumes perpendicular to the merging axis are observed, for example Abell 168 \citep{2004ApJ...610L..81H}, Abell 2146 \citep{2010MNRAS.406.1721R, 2012MNRAS.423..236R}.
\citet{2012MNRAS.423..236R} suggest that the structure in Abell 2146 is likely to be part of the remnant core of the primary subcluster, which was pushed forward and laterally by the collision.
This scenario can also be applied to this merging system where we observe two wings behind the outbound NW subcluster. 
 

However, the size of NE wing is considerable and is even larger than the NW conic subcluster. Thus, it may have a different origin. 
The optical analysis suggests that in addition to the $z=0.2687$ northern subcluster that is associated with the cone in X-ray and the $z=0.2684$ southern subcluster that is associated with the southern linear structure, a third subcluster at $z=0.2743$ is spatially overlaid with the northern subcluster \citep{2017ApJ...841....7B}.
The large NE wing may belong to the third subcluster and was displaced to the NE direction. To prove this scenario, we search for galaxies that around the NE wing tip in the redshift catalog of \citet{2017ApJ...841....7B}. We find several galaxies there, but all of them are at $z\sim0.270$. If the NE wing is associated with those galaxies, it is more likely to be from the either the $z=0.2687$ northern subcluster or the $z=0.2684$ southern subcluster. 

Therefore, we prefer the scenario of a displaced gas clump due to the head-on collision. The mystery of the remarkable size of the NE wing needs to be further investigated by numerical simulations with specific merging parameters. 

\section{Conclusion}\label{sect:conclusion}

We analyzed the deep \emph{Chandra} observations of merging galaxy cluster ZwCl 2341+0000. The cluster is  elongated in the NW-SE direction.  Meanwhile, we used \ac{GMRT} and \ac{JVLA} data to investigate the properties of the radio shocks. We summarize the main results as follows:
\begin{enumerate}
    \item The SE linear surface brightness edge is a kink in the new observations. It is the disrupted core of the southern subcluster. The shock front that is possibly responsible for the southern radio relic is revealed by the radio spectral index maps and is located at the SE edge of the relic component S2, which is $\sim160\arcsec$ (640 kpc) away from the SE surface brightness kink.
    \item At radio wavelengths, the Mach numbers of the southern and northern shocks are $2.2\pm0.1$ and $2.4\pm0.4,$ based on injection spectral indices and the assumption of \ac{DSA}. In X-rays, assuming the pre-shock ICM temperature follows a universal profile of relaxed clusters, we obtain a $3\sigma$ lower limit of the southern shock Mach number of 1.6.
    \item The NW subcluster has a cone shape with two linear $\sim400$ kpc cold fronts on the two sides. 
    We searched for analogies of the cone-shaped subcluster in numerical simulations and found that the cone is a certain stage of an outbound subcluster in the head-on merger, which happens between a bullet-like morphology and a slingshot phase. This stage is short-lived, and no other example has been discovered to date. 
    \item Behind the NW conic subcluster, there are two wing structures stretching out in the NE and western directions. The NE wing is broad and has a low temperature and entropy at the tip. The redshift of galaxies around the wingtip suggests that this structure is unlikely to be associated with the third subcluster discovered in the previous optical analysis. This may thus represent an unexpectedly large gas plume expanding perpendicular to the merger axis.
    \item A 400 kpc cool ($\sim4$ keV) gas bridge connects the apex of the NW cone and the NE wing. The low temperature suggests that this is a stripped gas trail from the NW subcluster. Why it is split from the subcluster is unclear. It could have evolved from a KHI eddy.
    \item There are two surface brightness bays located in the SW periphery. The southern bay and the ridge between the two bays have a $>10$ keV temperature, while the northern bay has a $3.3\pm0.7$ keV temperature. There is no pressure equilibrium between the ridge and the northern bay. The physical origin of these features is still unclear.
\end{enumerate}

In summary, ZwCl 2341+0000 is an exciting galaxy cluster merger that exhibits several unique morphology features. It is likely in a short-lived phase that is rarely observed and offers an example of the complex transition between a bullet-like morphology and the development of a slingshot tail.

\begin{acknowledgements}
X.Z. acknowledges support from Chinese Scholarship Council (CSC). A.S. is supported by the Women In Science Excel (WISE) programme of the Netherlands Organisation for Scientific Research (NWO), and acknowledges the World Premier Research Center Initiative (WPI) and the Kavli IPMU for the continued hospitality. SRON Netherlands Institute for Space Research is supported financially by NWO. R.J.vW. acknowledges support from the ERC Starting Grant ClusterWeb 804208. C.S. and A.B. acknowledge support from ERC-Stg DRANOEL n. 714245. This research has made use of data obtained from the Chandra Data Archive and the Chandra Source Catalog, and software provided by the Chandra X-ray Center (CXC) in the application package CIAO. The National Radio Astronomy Observatory is a facility of the National Science Foundation operated under cooperative agreement by Associated Universities, Inc. We thank the staff of the GMRT who have made these observations possible. The GMRT is run by the National Centre for Radio Astrophysics of the Tata Institute of Fundamental Research. This work made use of data from the Galaxy Cluster Merger Catalog. This research made use of Astropy,\footnote{\url{http://www.astropy.org}} a community-developed core Python package for Astronomy \citep{2013A&A...558A..33A,2018AJ....156..123A}
\end{acknowledgements}

\bibpunct{(}{)}{;}{a}{}{,}
\bibliographystyle{aa}
\bibliography{zwcl2341}

\begin{appendix}
\section{JVLA B-configuration radio map}
\begin{figure}[p]
\begin{center}
\includegraphics[width=1\hsize]{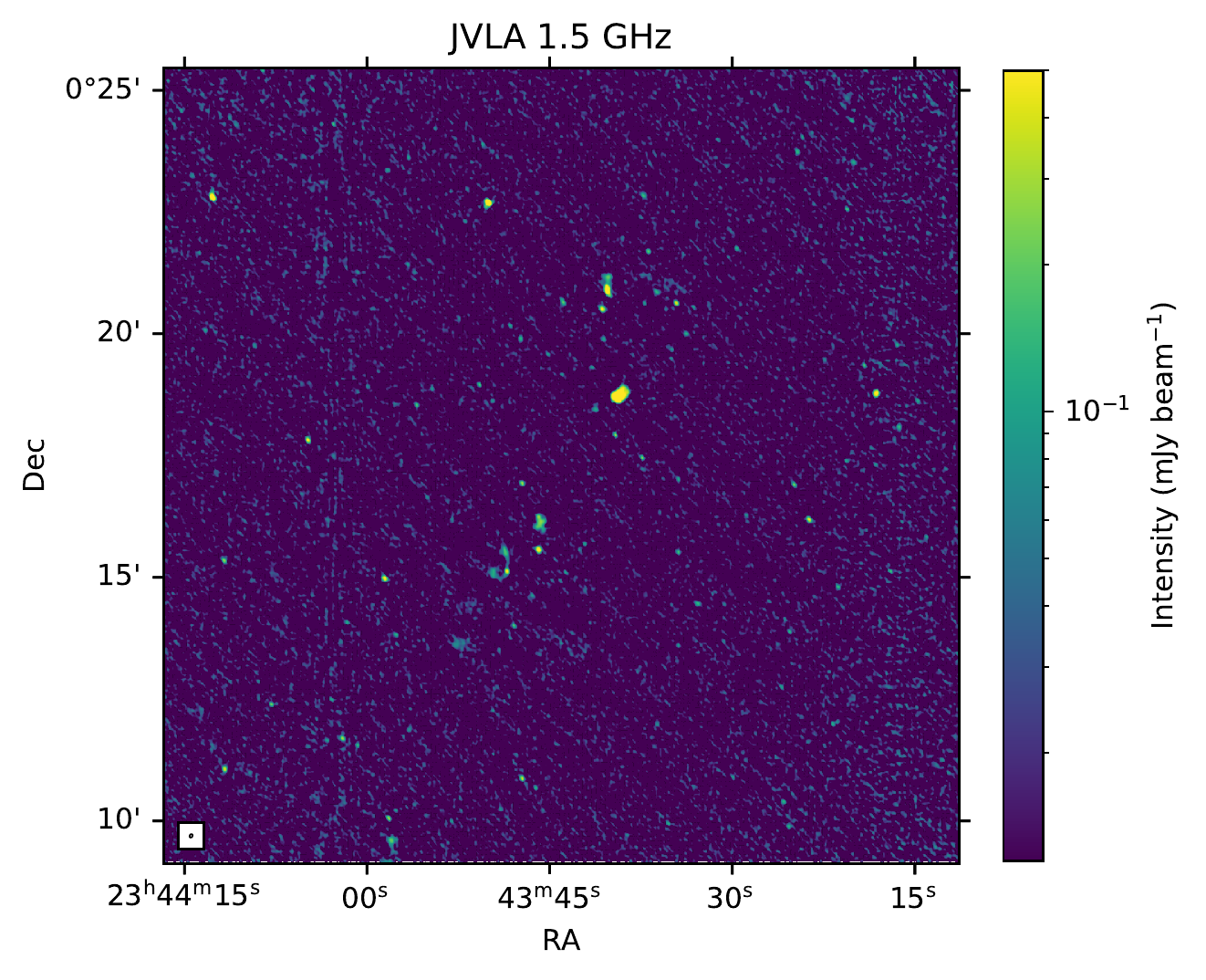}\\
\caption{
\ac{JVLA} 1.5 GHz B-configuration radio map of ZwCl 2341+0000.
}
\end{center}
\label{fig:vla-b}
\end{figure}

The \ac{JVLA} L-band B-configuration image is plotted in Fig. \ref{fig:vla-b}, where most of the relics are not visible.

\section{Radio spectral index uncertainty map}
\begin{figure*}[!]
\begin{center}
\begin{tabular}{ccc}
\includegraphics[height=2.6 in]{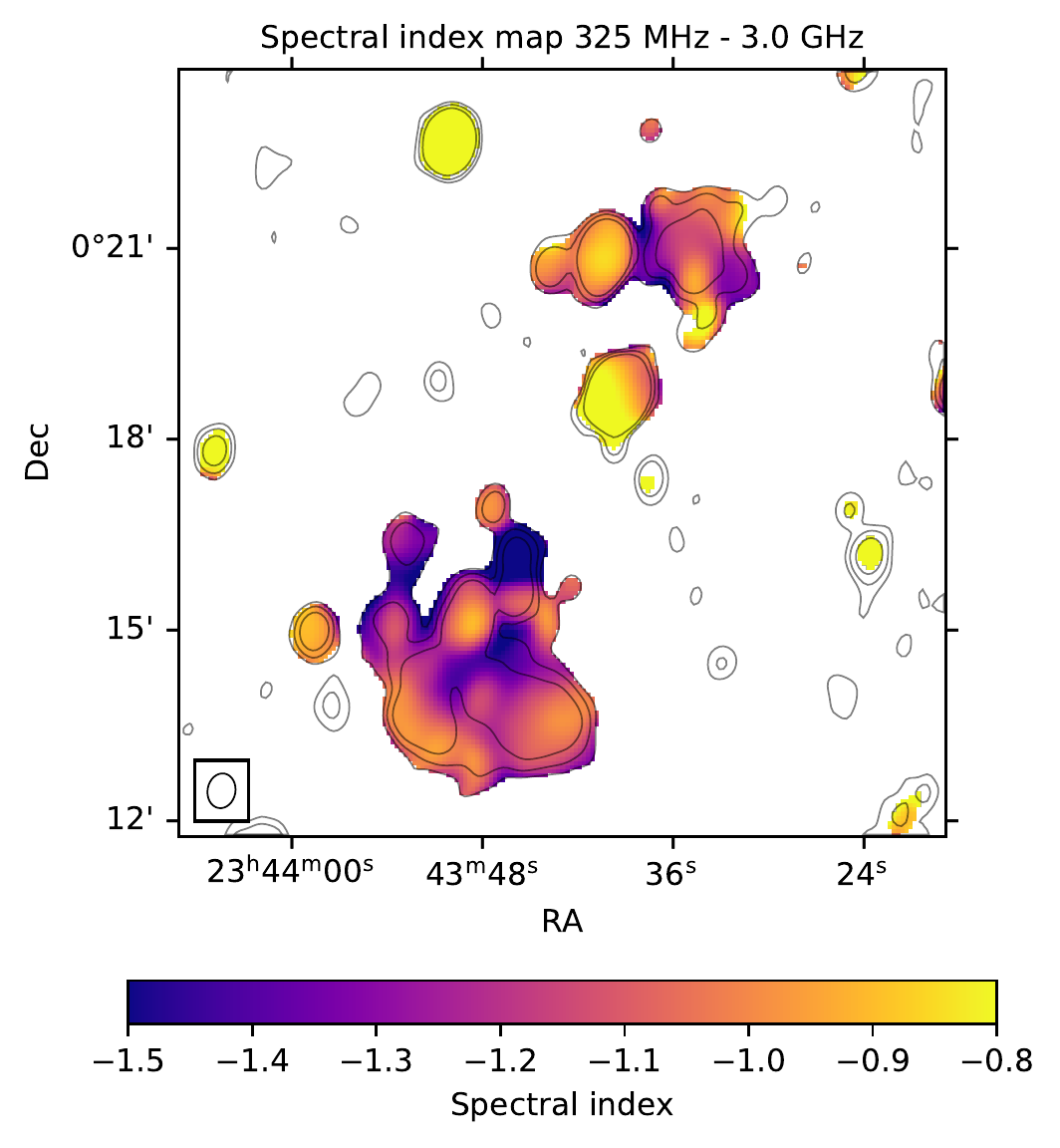}&
\includegraphics[height=2.6 in]{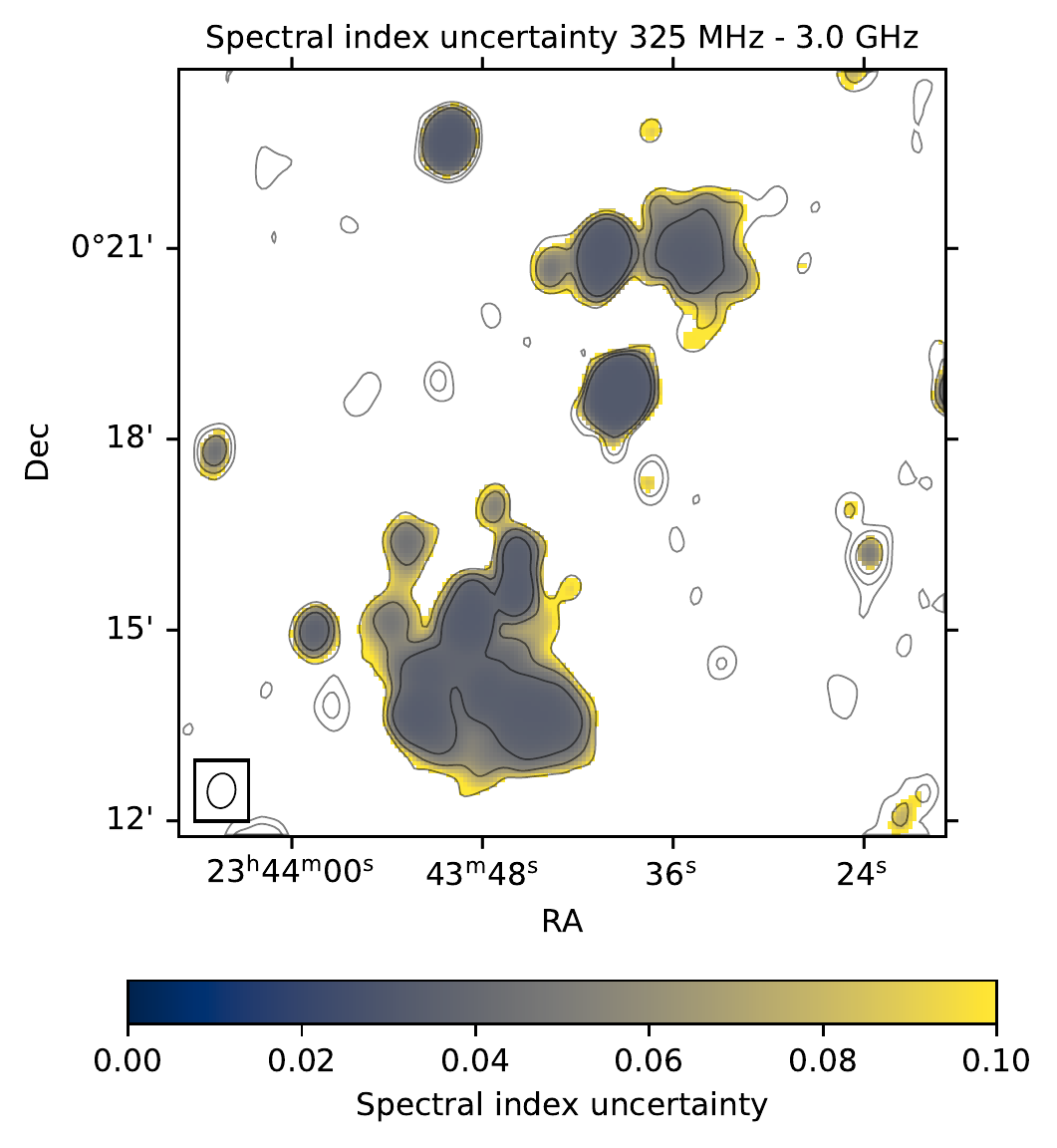}&
\includegraphics[height=2.6 in]{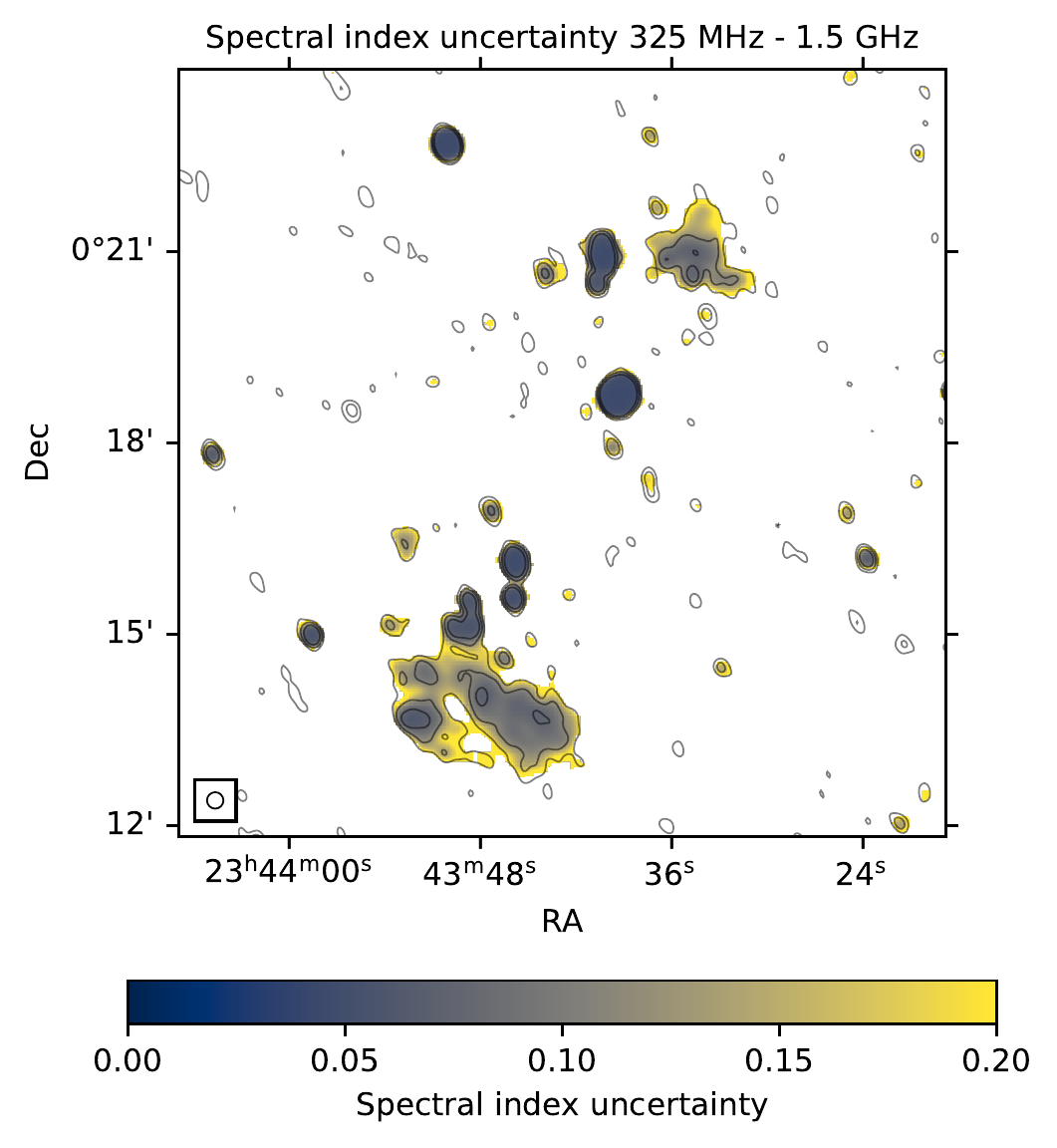}\\
\end{tabular}
\caption{\emph{Left and Middle:} Spectral index map and the uncertainty between 325 MHz and 3.0 GHz. \emph{Right:} Spectral index uncertainty map between 325 MHz and 1.5 GHz.}
\label{fig:spxerr}
\end{center}
\end{figure*}

The spectral index uncertainty map between 325 MHz and 1.5 GHz is plotted in Fig. \ref{fig:spxerr}. We also plot the spectral index map between 325 MHz and 3.0 GHz as well as its uncertainty map. 

\section{Temperature uncertainty map}\label{appendix:terrmap}
\begin{figure}[p]
\begin{center}
\includegraphics[width=0.99\hsize]{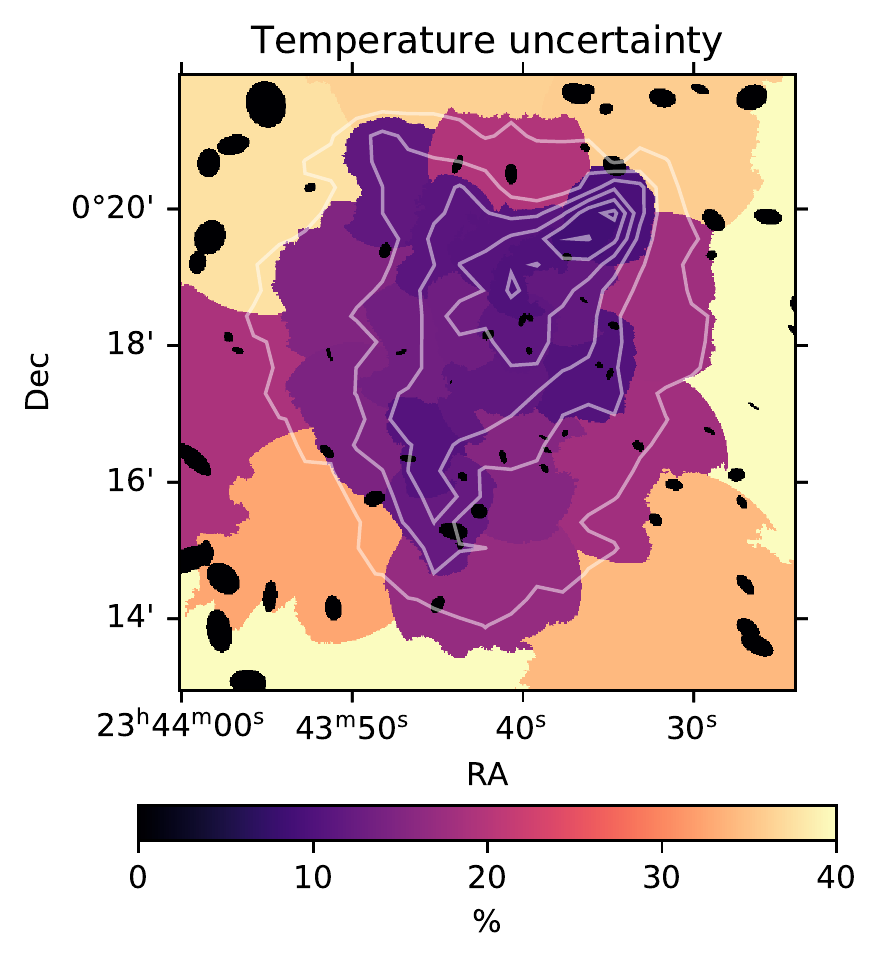}
\caption{Temperature uncertainty map of ZwCl 2341+0000 in units of percent, overlaid with the X-ray surface brightness contours. }
\label{fig:terrmap}
\end{center}
\end{figure}

The uncertainties of the pseudo pressure and entropy are dominated by the uncertainty of the temperature. The temperature uncertainty map (Fig. \ref{fig:terrmap}) shows that most regions of interests have an uncertainty less than $20\%$. Only the region HT3 has an uncertainty about $35\%$.

\section{Projected surface brightness profile for a cone-shaped cold front}\label{app:cdpl}
We assume the core remnant is in a right circular cone shape. For a given point in the space, the electron density can be expressed as a function of the distance to the apex $\rho$ and the angle with respect to the cone axis $\phi$ (i.e., $n_e=n_{e}\left(\rho,\phi\right))$. Similar to the broken power law model applied on a spheroid \citep{2009ApJ...704.1349O}, we define a broken power law density profile across a cone-shaped cold front,
\begin{equation}
n_e\left(\rho,\phi\right)=
\begin{cases}
n_0\left(\frac{\phi}{\phi_b}\right)^{-\alpha_1}\left(\frac{\rho}{\rho_0}\right)^{-\alpha_\rho}, & \phi\le\phi_b\\
\frac{n_0}{C}\left(\frac{\phi}{\phi_b}\right)^{-\alpha_2}\left(\frac{\rho}{\rho_0}\right)^{-\alpha_\rho},  & \phi_b<\phi\le\phi_\mathrm{max}
\end{cases}
,\end{equation}
where $n_0$ is the electron density normalization, $\phi_b$ the angle of the discontinuity, $C$ the density jump, $\alpha_1$ and $\alpha_2$ the power law index inside and outside the discontinuity, the term $\left(\rho/\rho_0\right)^{-\alpha_\rho}$ is the density variation along the axis, which is approximately negligible since the radial surface brightness is averaged. We assume the gas density is zero where $\phi>\phi_\mathrm{max}$.

The cone is projected to the plane of the sky. Assuming the axis is parallel to the sky plane, the integrated surface brightness at any point of the sky plane $(r, \theta)$ is
\begin{equation}
S\left(r,\theta\right)\propto\int_{-l_\mathrm{max}}^{l_\mathrm{max}} n_e^2 \mathrm{d}l, \ \theta\le\phi_\mathrm{max}
,\end{equation}
where $l_\mathrm{max}=r\times\left[1-\left(\cos\phi_\mathrm{max}/\cos{\theta}\right)^2\right]^{1/2}$ is the boundary of the integration. Therefore, the radially averaged surface brightness is 
\begin{equation}
S_\theta(\theta)=\frac{2}{r_\mathrm{max}^2-r_\mathrm{min}^2}\int_{r_\mathrm{min}}^{r_\mathrm{min}}S(r,\theta)r\mathrm{d}r,
\end{equation}
where $r_\mathrm{min}$ and $r_\mathrm{max}$ are the inner and outer radii of the extraction sector.

\end{appendix}

\end{document}